\documentclass[10pt,journal,letterpaper,compsoc]{IEEEtran}

\usepackage{amsmath}
\usepackage{amssymb}

\usepackage{subfigure}
\usepackage{algorithm}
\usepackage{algorithmic}
\usepackage[pagebackref=true,breaklinks=true,letterpaper=true,colorlinks,bookmarks=false]{hyperref}
\usepackage{xspace}
\usepackage{epsfig}
\usepackage{graphics}

\makeatletter
\DeclareRobustCommand\onedot{\futurelet\@let@token\@onedot}
\def\@onedot{\ifx\@let@token.\else.\null\fi\xspace}

\def\eg{\emph{e.g}\onedot} 
\def\ie{\emph{i.e}\onedot}

\def\etal{\emph{et al}\onedot}
\makeatother

\newtheorem{define}{Definition}

\def \x {\mathbf{x}}

\def \w {\mathbf{w}}

\makeatletter
\newcommand\figcaption{\def\@captype{figure}\caption}
\newcommand\tabcaption{\def\@captype{table}\caption}
\makeatother

\newcommand{\TripleFigureWidth}{165pt}
\newcommand{\TriQuaterFigureWidth}{145pt}

\newcommand{\nop}[1]{}

\begin{document}

\title{Density Sensitive Hashing}

\author{Yue Lin, Deng Cai,~\IEEEmembership{Member,~IEEE}, Cheng Li
\thanks{Y. Lin, D. Cai and C. Li are with the State Key Lab of CAD\&CG, College of Computer Science,
Zhejiang University, Hangzhou, Zhejiang, China, 310058.
Email: \texttt{linyue29@gmail.com},
\texttt{dengcai@cad.zju.edu.cn},\texttt{licheng@zju.edu.cn}.}
}

\IEEEcompsoctitleabstractindextext{%
\vspace{6mm}
\begin{abstract}
Nearest neighbors search is a fundamental problem in various research fields like machine learning, data mining and pattern recognition. Recently, hashing-based approaches, \eg, Locality Sensitive Hashing (LSH), are proved to be effective for scalable high dimensional nearest neighbors search.
Many hashing algorithms found their theoretic root in random projection. Since these algorithms generate the hash tables (projections) randomly, a large number of hash tables (\ie, long codewords) are required in order to achieve both high precision and recall. To address this limitation, we propose a novel hashing algorithm called {\em Density Sensitive Hashing} (DSH) in this paper. DSH can be regarded as an extension of LSH. By exploring the geometric structure of the data, DSH avoids the purely random projections selection and uses those projective functions which best agree with the distribution of the data.
Extensive experimental results on real-world data sets have shown that the proposed method achieves better performance compared to the state-of-the-art hashing approaches.
\end{abstract}

\begin{IEEEkeywords}
Locality Sensitive Hashing, Random Projection, Clustering.
\end{IEEEkeywords}}

\maketitle

\IEEEdisplaynotcompsoctitleabstractindextext

%
\IEEEpeerreviewmaketitle

\section{Introduction}
Nearest Neighbors (NN) search is a fundamental problem and has found applications in many data mining tasks \cite{dasgupta2011fast,gao2011continuous,he2010scalable}. A number of efficient algorithms, based on pre-built index structures (e.g. KD-tree~\cite{bentley1975multidimensional} and R-tree~\cite{arge2008priority}), have been proposed for nearest neighbors search. Unfortunately, these approaches perform worse than a linear scan when the dimensionality of the space is high~\cite{beyer1999nearest}.

Given the intrinsic difficulty of exact nearest neighbors search, many hashing algorithms are proposed for Approximate Nearest Neighbors (ANN) search~\cite{andoni2008near,charikar2002similarity,datar2004locality}. The key idea of these approaches is to generate binary codewords for high dimensional data points that preserve the similarity between them. Roughly, these hashing methods can be divided into two groups, the random projection based methods and the learning based methods.

Many hashing algorithms are based on the random projection, which has been proved to be an effective method to preserve pairwise distances for data points. One of the most popular methods is Locality Sensitive Hashing(LSH) \cite{andoni2008near,charikar2002similarity,datar2004locality,gionis1999similarity}. Given a database with $n$ samples, LSH makes no prior assumption about the data distribution and offers probabilistic guarantees of retrieving items within $(1+\epsilon)$ times the optimal similarity, with query times that are sub-linear with respect to $n$ \cite{kulis2009kernelized,norouzi2011minimal}. However, according to the Jonson Lindenstrauss Theorem~\cite{johnson1984extensions}, LSH needs $O(\ln n/\epsilon^2)$ random projections to preserve the pairwise distances, where $\epsilon$ is the relative error. Therefore, in order to increase the probability that similar objects are mapped to similar hash codes, LSH needs to use many random vectors to generate the hash tables (a long codeword), leading to a large storage space and a high computational cost.

Aiming at making full use of the structure of the data, many learning-based hashing algorithms \cite{brandt2010transform,jegou2011product,jiang2011semi,salakhutdinov2009semantic,jun2010semi,weiss2008spectral,zhang2010self} are proposed. Most of these algorithms exploit the spectral properties of the data affinity (\ie, item-item similarity) matrix for binary coding. Despite the success of these approaches for relatively small codes, they often fail to make significant improvement as the code length increases~\cite{joly2011random}.

In this paper, we propose a novel hashing algorithm called \emph{Density Sensitive Hashing} (DSH) for effective high dimensional nearest neighbors search. Our algorithm can be regarded as an extension of LSH. Different from all the existing random projection based hashing methods, DSH tries to utilize
the geometric structure of the data to guide the projections (hash tables) selection. Specifically, DSH uses $k$-means to roughly partition the data set into $k$ groups. Then for each pair of adjacent groups, DSH generates one projection vector which can well split the two corresponding groups. From all the generated projections, DSH select the final ones according to the maximum entropy principle, in order to maximize the information provided by each bit. Experimental results show the superior performance of the proposed Density Sensitive Hashing algorithm over the existing state-of-the-art approaches.

The remainder of this paper is organized as follows. We introduce the background and review the related work in Section 2. Our Density Sensitive Hashing algorithm is presented in Section 3. Section 4 gives the experimental results that compared our algorithm with the state-of-the-art hashing methods on three real world large scale data sets. Conclusion remarks are provided in Section 5.

\section{Background and Related Work}

The generic hashing problem is the following.
Given $n$ data points $\mathbf{X} = [\x_1,\cdots,\x_n] \in \mathbb{R}^{d \times n}$, find $L$ hash functions
to map a data point $\mathbf{x}$ to a $L$-bits hash code $$H(x)=[h_1(\mathbf{x}),h_2(\mathbf{x}),...,h_L(\mathbf{x})],$$ where $h_l(\mathbf{x}) \in \{0, 1\}$ is the $l$-th hash function. For the linear projection-based hashing, we have~\cite{wang2010sequential}
\begin{equation}\label{eq:1}
h_{l}(\mathbf{x})=sgn(F(\mathbf{w}_{l}^{T}\mathbf{x}+t_{l}))
\end{equation}
where $\mathbf{w}_{l}$ is the projection vector and $t_{l}$ is the intercept. Different hashing algorithms aim at finding different $F$, $\w_l$ and $t_l$ with respect to different objective functions.

One of the most popular hashing algorithms is Locality Sensitive Hashing (LSH) \cite{andoni2008near,charikar2002similarity,datar2004locality,gionis1999similarity}. LSH is fundamentally based on the random projection and uses
randomly generated $\w_l$. The $F$ in LSH is an identity function and $t_{l}=0$ for mean thresholding\footnote{Without loss of generality, we assume that all the data points are centralized to have zero mean.}. Thus, for LSH, we have
\begin{equation}
h_{l}(\mathbf{x}) = \left\{ \begin{array}{ll}
1 & \textrm{if $\mathbf{w}_{l}^{T}\mathbf{x} \ge 0$}\\
0 & \textrm{otherwise}
\end{array} \right.
\end{equation}
where $\mathbf{w}_{l}$ is a vector generated from a zero-mean multivariate Gaussian $\mathcal{N}(0, \mathbf{I})$ of the same dimension as the input $\mathbf{x}$. From the geometric point of view, the $\mathbf{w}_{l}$ defines a hyperplane. The points on different sides of the hyperplane have the opposite labels. Using this hash function, two points' hash bits match with the probability proportional to the cosine of the angle between them~\cite{charikar2002similarity}.
Specifically, for any two points $\mathbf{x}_{i},\mathbf{x}_{j} \in \mathbb{R}^{d}$, we have~\cite{kulis2009kernelized}:
\begin{equation}
Pr[h_{l}(\mathbf{x}_{i})=h_{l}(\mathbf{x}_{j})]=1-\frac{1}{\pi}\cos^{-1}(\frac{\mathbf{x}_{i}^{T}\mathbf{x}_{j}}{\big\|\mathbf{x}_{i}\big\|\big\|\mathbf{x}_{j}\big\|})
\end{equation}
Based on this nice property, LSH have the
probabilistic guarantees of retrieving items within $(1+\epsilon)$ times the optimal similarity, with query times that are sub-linear with respect to $n$ \cite{kulis2009kernelized,norouzi2011minimal}.

Empirical studies~\cite{andoni2008near} showed that the LSH is significantly more efficient than the methods based on hierarchical tree decomposition. It has been successfully used in various applications in data mining~\cite{dasgupta2011fast,he2010scalable}, computer vision~\cite{jun2010semi,wang2006annosearch} and database~\cite{mohammad2004voronoi,korn1996fast}. There are many extensions for LSH~\cite{joly2008a,kulis2009kernelized,lv2007multi,panigrahy2006entropy}.
Entropy based LSH~\cite{panigrahy2006entropy} and Multi-Probe LSH~\cite{lv2007multi,joly2008a} are proposed to reduce the space requirement in LSH but need much longer time to deal with the query. The original LSH methods cannot apply for high-dimensional kernelized data when the underlying feature embedding for the kernel is unknown. To address this limitation, Kernelized Locality Sensitive Hashing is introduced in~\cite{kulis2009kernelized}. It suggests to approximate a normal distribution in the kernel space using only kernel comparisons~\cite{joly2011random}. In addition, the Shift Invariant Kernels Hashing~\cite{raginsky2009locality}, which is a distribution-free method based on the random features mapping for shift-invariant kernels, is also proposed recently. This method has theoretical convergence guarantees and performs well for relatively large code sizes~\cite{gong2011iterative}.
All these methods are fundamentally based on the random projection.
According to the Jonson Lindenstrauss Theorem~\cite{johnson1984extensions}, $O(\ln n/\epsilon^2)$ projective vectors are needed to preserve the pairwise distances of a database with size $n$ for the random projection, where $\epsilon$ is the relative error. Therefore, in order to increase the probability that similar objects are mapped to similar hash codes, these random projection based hashing methods need to use many random vectors to generate the hash tables (a long codeword), leading to a large storage space and a high computational cost.

To address the above limitation, many learning-based hashing methods~\cite{kulis2010learning,brandt2010transform,gong2011iterative,jegou2011product,jiang2011semi,joly2011random,li2011hashing,wei2011hashing,mu2010weakly,norouzi2011minimal,pauleve2010locality,salakhutdinov2009semantic,jun2010semi,weiss2008spectral,xu2011complementary,zhang2010laplacian,zhang2010self} are proposed. PCA Hashing \cite{wang2006annosearch} might be the simplest one. It chooses $\mathbf{w}_l$ in Eq.(\ref{eq:1}) to be the principal directions of data. Many other algorithms \cite{wei2011hashing,wang2010sequential,weiss2008spectral,zhang2010self} exploit the spectral properties of the data affinity (\ie, item-item similarity) matrix for binary coding. The spectral analysis of the data affinity matrix is usually time consuming \cite{Cai09SR-Thesis}. To avoid the high computational cost, Weiss \etal \cite{weiss2008spectral} made a strong assumption that data is uniformly
distributed and proposed a Spectral Hashing method (SpH). The assumption in SpH leads to a simple analytical eigenfunction solution of 1-D Laplacians, but the geometric
structure of the original data is almost ignored, leading to a suboptimal performance. Anchor Graph Hashing (AGH) \cite{wei2011hashing} is a recently proposed method to overcome this shortcoming. AGH generates $k$ anchor points from the data and represents all the data points by sparse linear combinations of the anchors. In this way, the spectral analysis of the data affinity can be efficiently performed. Some other learning based hashing methods include Semantic Hashing \cite{salakhutdinov2009semantic} which uses the stacked Restricted Boltzmann Machine (RBM) to generate the compact binary codes; Semi-supervised Sequential Projection Hashing (S3PH) \cite{wang2010sequential} which can incorporate  supervision information. Despite the success of these learning based hashing approaches for relatively small codes, they often fail to make significant improvement as the code length increases~\cite{joly2011random}.

\begin{figure*}[t]
\begin{center}
\subfigure[Locality Sensitive Hashing~\cite{charikar2002similarity}]{\includegraphics[width=\TripleFigureWidth]{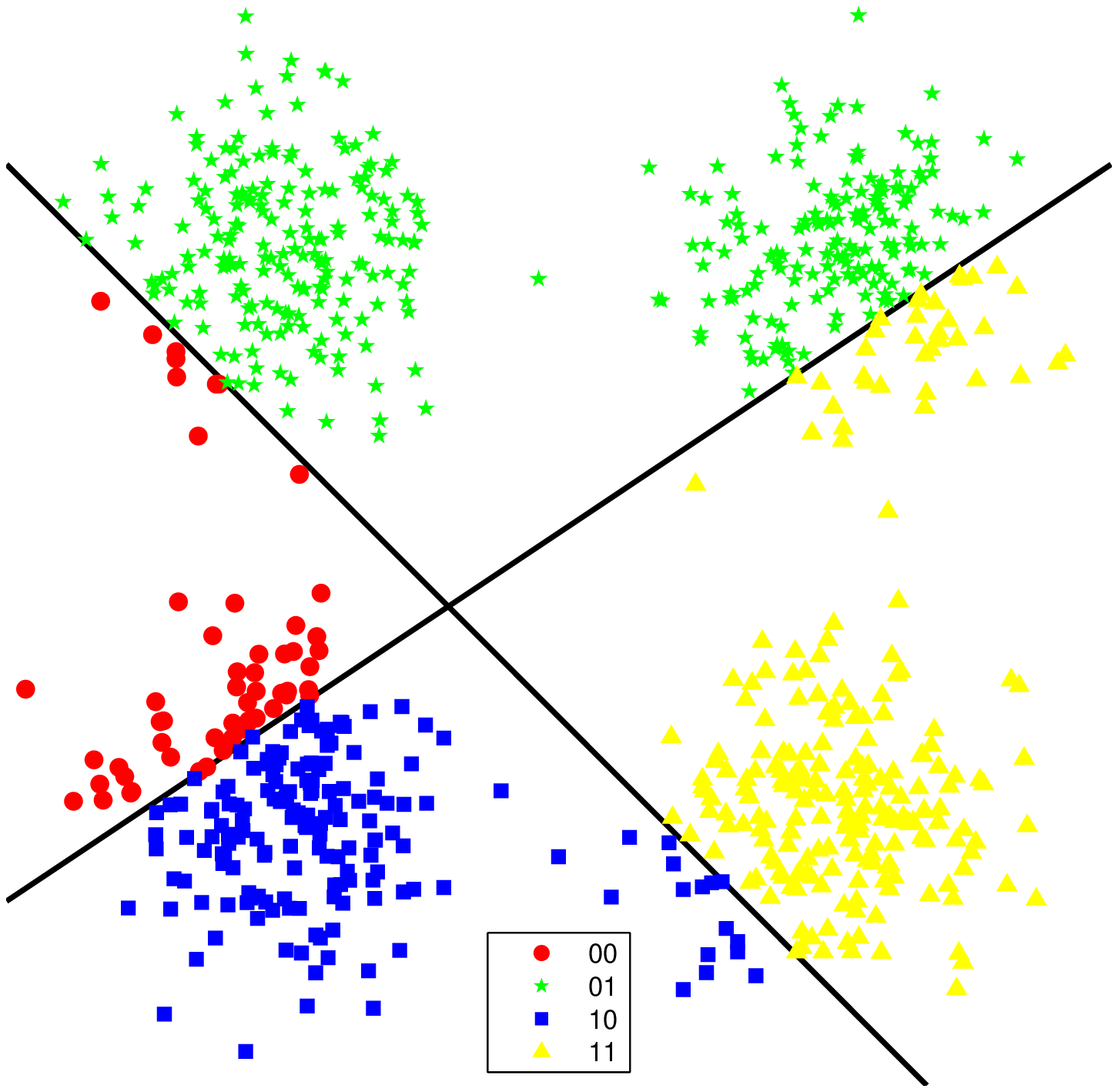}}
\subfigure[PCA Hashing~\cite{wang2006annosearch}]{\includegraphics[width=\TripleFigureWidth]{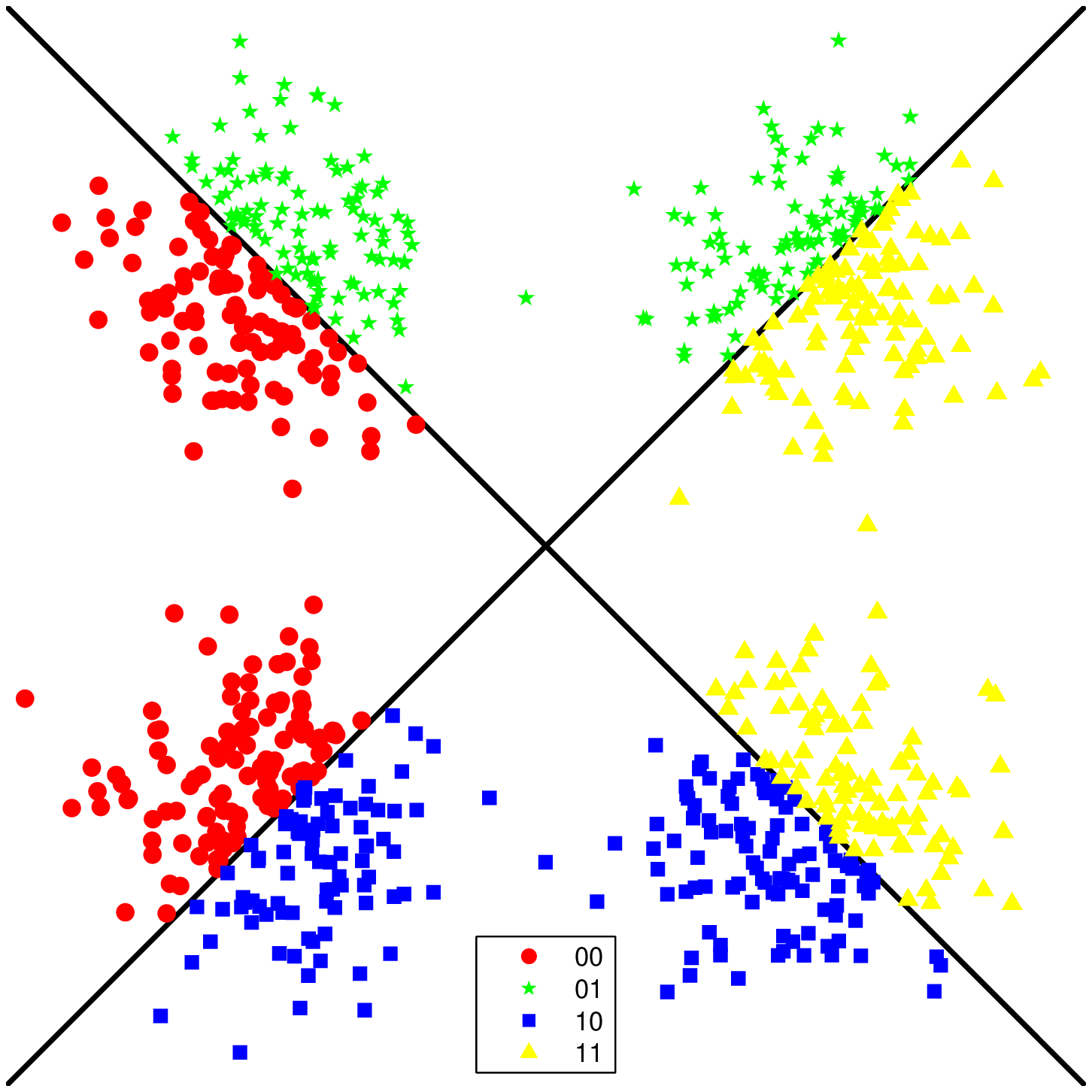}}
\subfigure[Density Sensitive Hashing]{\includegraphics[width=\TripleFigureWidth]{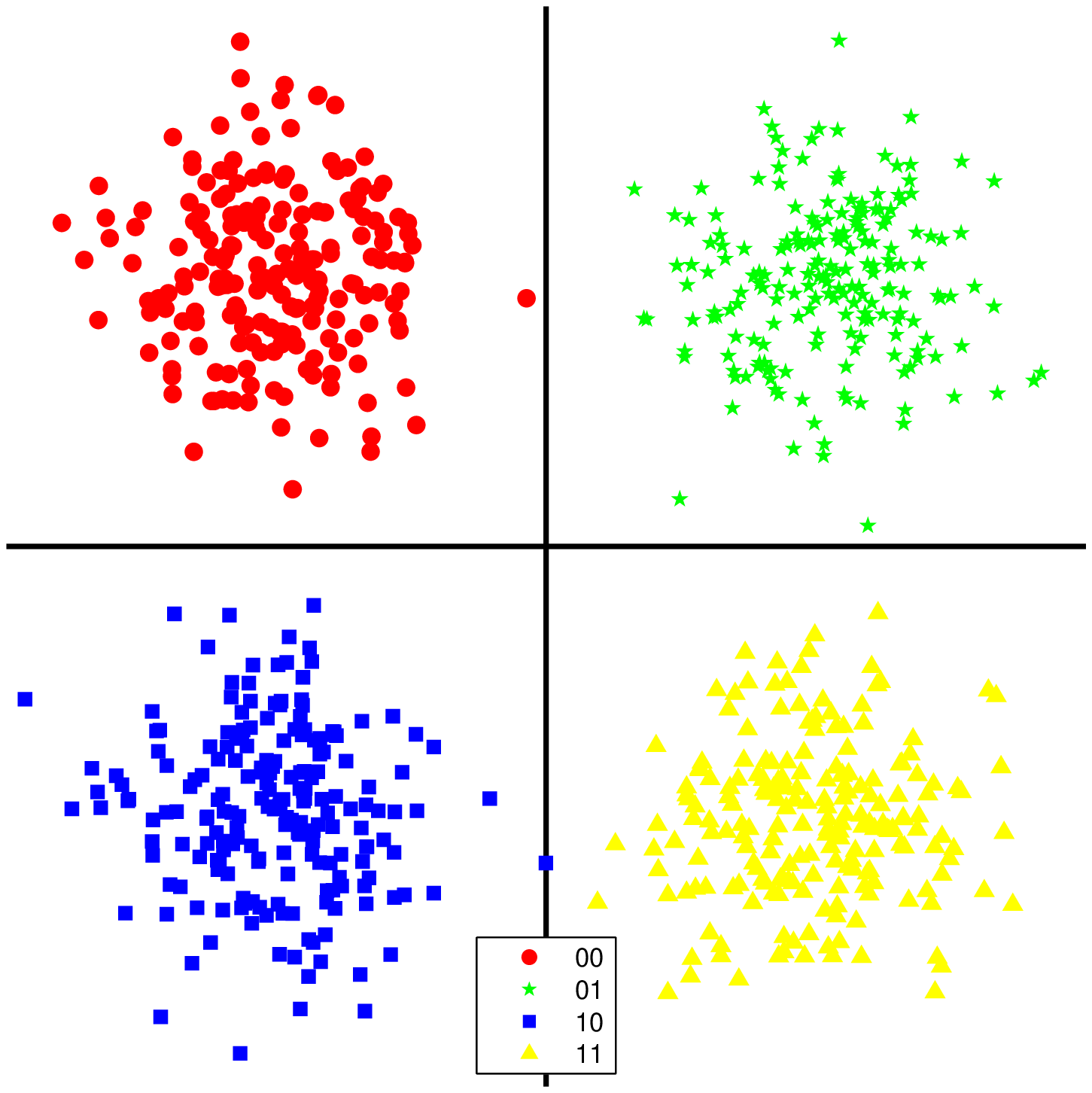}}\hfill
\caption{Illustration of different hashing methods on a toy data set. There are four Gaussians in a two dimensional plane and one is asked to encode the data using  two-bits hash codes.   (a) LSH generates the projections randomly and it is very possible that the data points from the same Gaussian will be encoded by different hash codes. (b) PCA Hashing uses the principle directions of the data as the projective vectors. In our example, all the four Gaussians are split and PCA Hashing generates an unsatisfactory coding. (c) Considering the geometric structure of the data (density of the data), our DSH
generates perfect binary codes for this toy example.} \label{fig:101}
\end{center}
\end{figure*}

\section{Density Sensitive Hashing}

In this section, we give the detailed description on our proposed {\em Density Sensitive Hashing} (DSH) which aims at overcoming the disadvantages of both random projection based and learning based hashing approaches. To guarantee the performance will increase as the code length increases, DSH adopts the similar framework as LSH. Different from LSH which generates the projections randomly, DSH uses the geometric structure of the data to guide the selection of the projections.

Figure~\ref{fig:101} presents a toy example to illustrate the basic idea of our approach. There are four Gaussians in a two dimensional plane and one is asked to encode the data using  two-bits hash codes.   LSH \cite{charikar2002similarity} generates the projections randomly and it is very possible that the data points from the same Gaussian will be encoded by different hash codes. PCA Hashing \cite{wang2006annosearch} uses the principle directions of the data as the projective vectors. In our example, all the four Gaussians are split and PCA Hashing generates an unsatisfactory coding. Considering the geometric structure of the data (density of the data), our DSH
generates perfect binary codes for this toy example. The detailed procedure of DSH will be provided in the following subsections.

\subsection{Minimum Distortion Quantization}

The first step of DSH is quantization of the data. Recently, Paulev{\'e} \etal \cite{pauleve2010locality} show that
a quantized preprocess for the data points can significantly improve the performance of the nearest neighbors search. Motivated by this result, we use the $k$-means algorithm, one of the most popular quantization approaches, to partition the $n$ points into $k$ ($k < n$) groups.

Let $\mathcal{S}=\{\mathcal{S}_{1},\cdots,\mathcal{S}_{k}\}$ denote a given quantization result. The \emph{distortion}, also known as the \emph{Sum of Squared Error} (SSE), can be used to measure the quality of the given quantization:
\begin{equation}
SSE=\sum_{p=1}^{k}\sum_{\mathbf{x} \in \mathcal{S}_{p}}\|\mathbf{x}-\mathbf{\mu}_{p}\|^{2}
\end{equation}
The $\mathbf{\mu}_p$ is the representative point of the $p$-th group $\mathcal{S}_{p}$. By noticing
\begin{eqnarray*}
\frac{\partial SSE}{\partial \mathbf{\mu}_{i}}&=&\frac{\partial}{\partial \mathbf{\mu}_{i}}\sum_{p=1}^{k}\sum_{\mathbf{x} \in \mathcal{S}_{p}}\|\mathbf{x}-\mathbf{\mu}_{p}\|^{2}\\
&=&\sum_{p=1}^{k}\sum_{\mathbf{x} \in \mathcal{S}_{p}}\frac{\partial}{\mathbf{\mu}_{i}}\|\mathbf{x}-\mathbf{\mu}_{p}\|^{2}\\
&=&\sum_{\mathbf{x} \in \mathcal{S}_{i}}2(\mathbf{x}-\mathbf{\mu}_{i})=0,\;\;\;i=1,\cdots,k,
\end{eqnarray*}
we have:
\begin{equation}
\mathbf{\mu}_{i}=\frac{1}{|\mathcal{S}_{i}|}\sum_{\mathbf{x} \in \mathcal{S}_{i}}\mathbf{x},\;\;\;i=1,2,...,k
\end{equation}
It indicates that in order to minimize the distortion, we can choose the center point as the representative point for each group.

There are two points that needed to be highlighted for the $k$-means quantization in our approach:
\begin{enumerate}
  \item In large scale applications, it can be time consuming to wait the $k$-means converges. Naturally, we can stop the $k$-means after $p$ iterations, where $p$ is a parameter. We found that a small number of $p$ is usually enough (usually 5). This will be discussed in the our experiments.

  \item In real applications, we do not know which is the best group number $k$. It seems that the bigger the $k$, the better performance we will get. It is simply because the quantization will have smaller error with a large number of groups. However, a large number of groups could lead to high computational cost in the quantization step. As will be described in the next subsection, the number of groups decides the maximum code length DSH can generate. Thus, we set
      \begin{equation}
        k = \alpha L
      \end{equation}
      where $L$ is the code length and $\alpha$ is a parameter.
\end{enumerate}

\subsection{Density Sensitive Projections Generation}

Now we have the quantization result denoted by $k$ groups $\mathcal{S}_{1},\cdots,\mathcal{S}_{k}$ and the $i$-th group has
the center $\mu_i$. Instead of generating projections randomly as LSH does, our DSH tries to use this quantization result to guide the projections generating process.

We define the $r$-nearest neighbors matrix $\textbf{W}$ of the groups as follows:
\begin{define}\label{def:1}
{\bf $r$-Nearest Neighbors Matrix $\textbf{W}$} of the groups.
\begin{equation}
\textbf{W}_{ij}=\left\{
         \begin{array}{ll}
           1, & \hbox{if $\mu_i \in N_r(\mu_j)$ or $\mu_j \in N_r(\mu_i)$} \\
           0, & \hbox{otherwise}
         \end{array}
       \right.
\end{equation}
where $N_r(\mu_i)$ denotes the set of $r$ nearest neighbors of
$\mu_i$.
\end{define}
With this definition, we can then define $r$-adjacent groups:
\begin{define}\label{def:2}
{\bf $r$-Adjacent Groups:} Group $\mathcal{S}_{i}$ and group $\mathcal{S}_{j}$ are called $r$-adjacent groups if and only if
$\textbf{W}_{ij}=1$.
\end{define}
Instead of picking a random projection, it is more natural to pick those projections which can well separate two adjacent groups.

For each pair of adjacent groups $\mathcal{S}_{i}$ and $\mathcal{S}_{j}$, DSH uses the median plane between the centers of adjacent groups as the hyperplane to separate points. The median plane is defined as follows:
\begin{equation}\label{eq:9}
(\x-\frac{\mathbf{\mu}_{i}+\mathbf{\mu}_{j}}{2})^{T}(\mathbf{\mu}_{i}-\mathbf{\mu}_{j})=0
\end{equation}
One can easily verify that the hash function associated with this plane is defined as follows:
\begin{equation} \label{eq:10}
h(\x) = \left\{ \begin{array}{ll}
1 & \textrm{if $\w^{T} \x \ge t$}\\
0 & \textrm{otherwise}
\end{array} \right.
\end{equation}
where
\begin{equation} \label{eq:105}
\w=\mu_1 - \mu_2,\;\;\;t=(\frac{\mu_1 + \mu_2}{2})^{T}(\mu_1 - \mu_2)
\end{equation}

\subsection{Entropy Based Projections Selection}

Given $k$ groups, the previous step can generate around $\frac{1}{2}kr$ projections. Since $k=\alpha L$, our DSH generates $\frac{1}{2}\alpha rL$ projections so far. Each projection will lead to one bit in the code and the usual setting of the parameters $\alpha$, $r$ will make $\frac{1}{2}\alpha rL > L$. Thus, our DSH needs a projections selection step which aims at selecting $L$ projections from the candidate set containing $\frac{1}{2}\alpha rL$ projections.

From the information theoretic point of view, a "good" binary codes should maximize the information/entropy provided by each bit~\cite{zhang2010self}. Using maximum entropy principle, a binary bit that gives balanced partitioning of the data points provides maximum information~\cite{jun2010semi}. Thus, we compute the entropy of each candidate projection and select the projections which can split the data most equally.

Assume we have $m$ candidate projections $\w_{1},\w_{2},...,\w_{m}$. For each projection, the data points are separated into two sets and labeled with opposite bit. We denote these two partitions as $\mathcal{T}_{i0}$ and $\mathcal{T}_{i1}$, respectively. The entropy $\mathcal{\delta}_{i}$ with respect to the projection $\w_i$ can be computed as:
\begin{equation}\label{eq:11}
\mathcal{\delta}_{i}=-P_{i0} \log P_{i0}-P_{i1} \log P_{i1}
\end{equation}
where:
\begin{eqnarray}\label{eq:12}
P_{i0}=\frac{|\mathcal{T}_{i0}|}{|\mathcal{T}_{i0}|+|\mathcal{T}_{i1}|},\;\;P_{i1}=\frac{|\mathcal{T}_{i1}|}{|\mathcal{T}_{i0}|+|\mathcal{T}_{i1}|}
\end{eqnarray}
In practice, the database can be very large and computing the entropy of each projection with respect to the entire database is time consuming. Thus, we estimate the entropy simply by using the group centers. For group center $\mu_i$, we assign a weight $\nu_{i}$ based on the size of the group.
\begin{equation}\label{eq:13}
\nu_{i}=\frac{|\mathcal{S}_{i}|}{\sum_{p=1}^{k}|\mathcal{S}_{p}|}
\end{equation}
We denote the two sets of group centers as $\mathcal{C}_{i0}$ and $\mathcal{C}_{i1}$.
Then $P_{i0}$ and $P_{i1}$ can be computed as:
\begin{equation}\label{eq:14}
P_{i0}=\sum_{s \in \mathcal{C}_{i0}}\nu_{s},\;\;\;P_{i1}=\sum_{t \in \mathcal{C}_{i1}}\nu_{t}
\end{equation}
This simplification significantly reduces the time cost on the entropy calculation.

After obtaining the entry $\mathcal{\delta}_{i}$ for each $\w_i$, we sort them in descending order and use the top $L$ projections for creating the $L$-bit binary codes, according to Eq.(\ref{eq:10}). The overall procedure of our DSH algorithm is summarized in Alg.~\ref{alg:1}.

\begin{algorithm}[t]
   \caption{Density Sensitive Hashing}
   \label{alg:1}
\begin{algorithmic}[1]
\REQUIRE ~~\\
$n$ training samples $ \textbf{x}_1, \textbf{x}_2, \ldots, \textbf{x}_n \in \mathbb{R}^d$; \\
$L$: the number of bits for hashing codes;\\
$\alpha$: the parameter controlling the groups number;\\
$p$: the number of iterations in the $k$-means;\\
$r$: the parameter for $r$-adjacent groups\\
   \STATE Use $k$-means with $p$ iterations to generate $\alpha L$ groups, with centers $\mathcal{\mu}_{1},\cdots,\mathcal{\mu}_{\alpha L}$. \label{st:1}

   \STATE Generate the list of all $r$-adjacent groups based on the definition (\ref{def:1}) and (\ref{def:2}). \label{st:2}

   \STATE For each pair of adjacent groups, use Eq.(\ref{eq:105}) to generate the projection $\w$ and intercept $t$. \label{st:3}

   \STATE Calculate the entropy of all the candidate projections using the weighted center points based on Eq.(\ref{eq:11}) and Eq.(\ref{eq:14})\label{st:4}

   \STATE Sort the entropy values in descending order and use the top $L$ projections to create binary codes according to Eq.(\ref{eq:10}). \label{st:5}

\ENSURE ~~\\
   The model: $\{\w_i,t_i\}_{i=1}^L$\\
   Binary hashing codes for the training samples: $Y \in \{0, 1\}^{n\times L}$
   \end{algorithmic}
\end{algorithm}

\begin{figure*}[t!]
\begin{center}
\subfigure[GIST1M]{
\includegraphics[width=0.65\columnwidth]{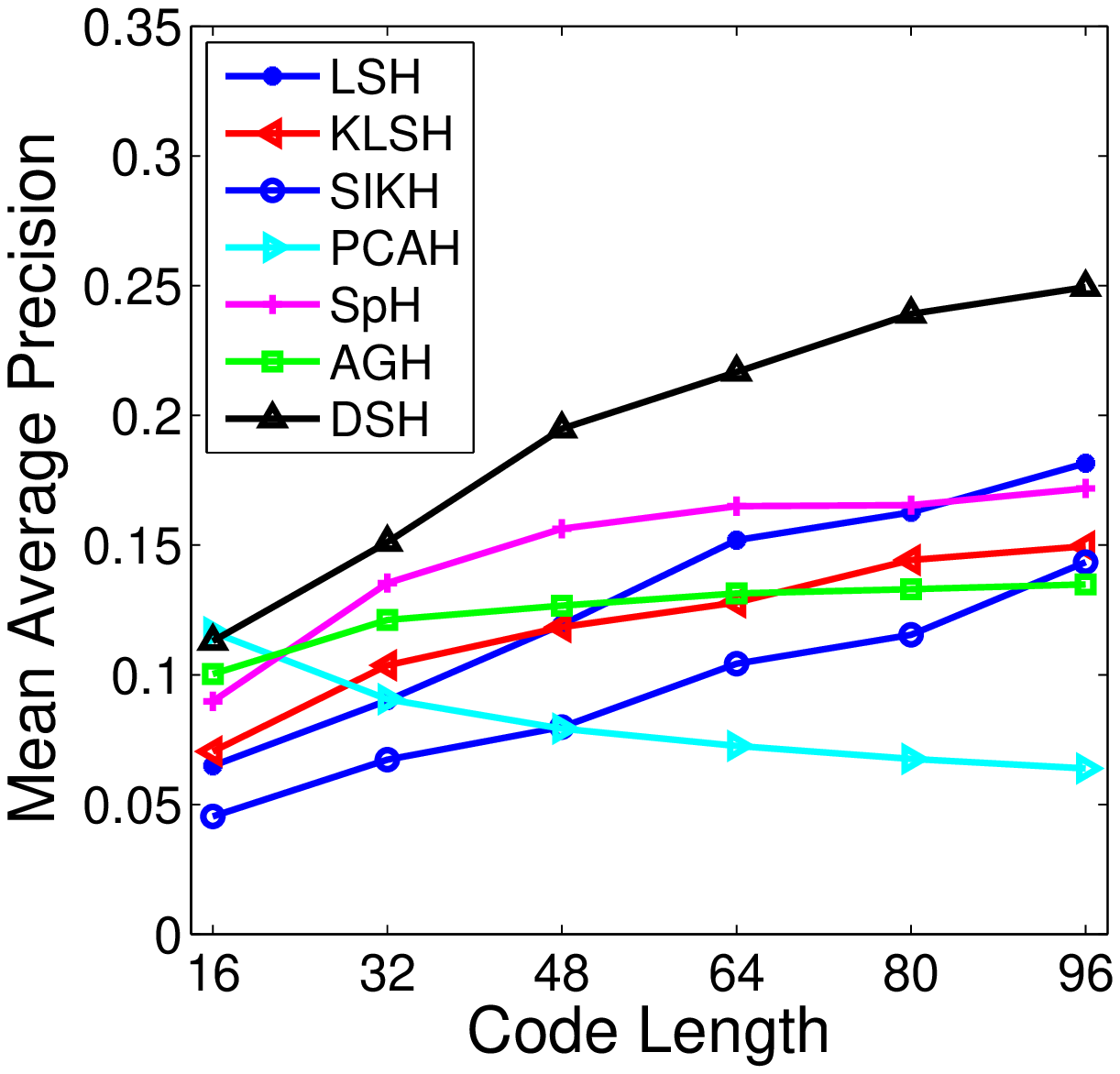}}
\subfigure[Flickr1M]{
\includegraphics[width=0.65\columnwidth]{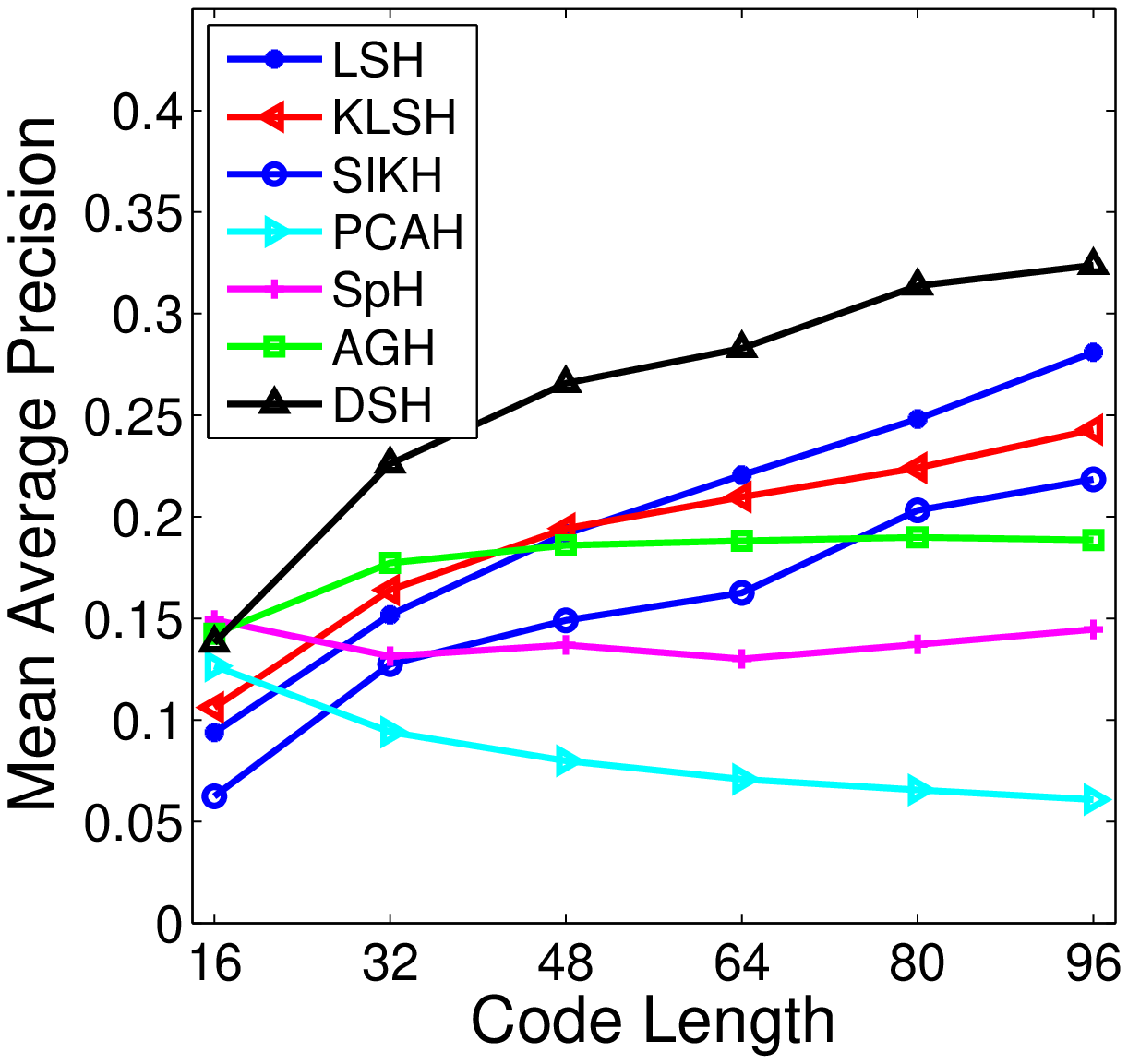}}
\subfigure[SIFT1M]{
\includegraphics[width=0.64\columnwidth]{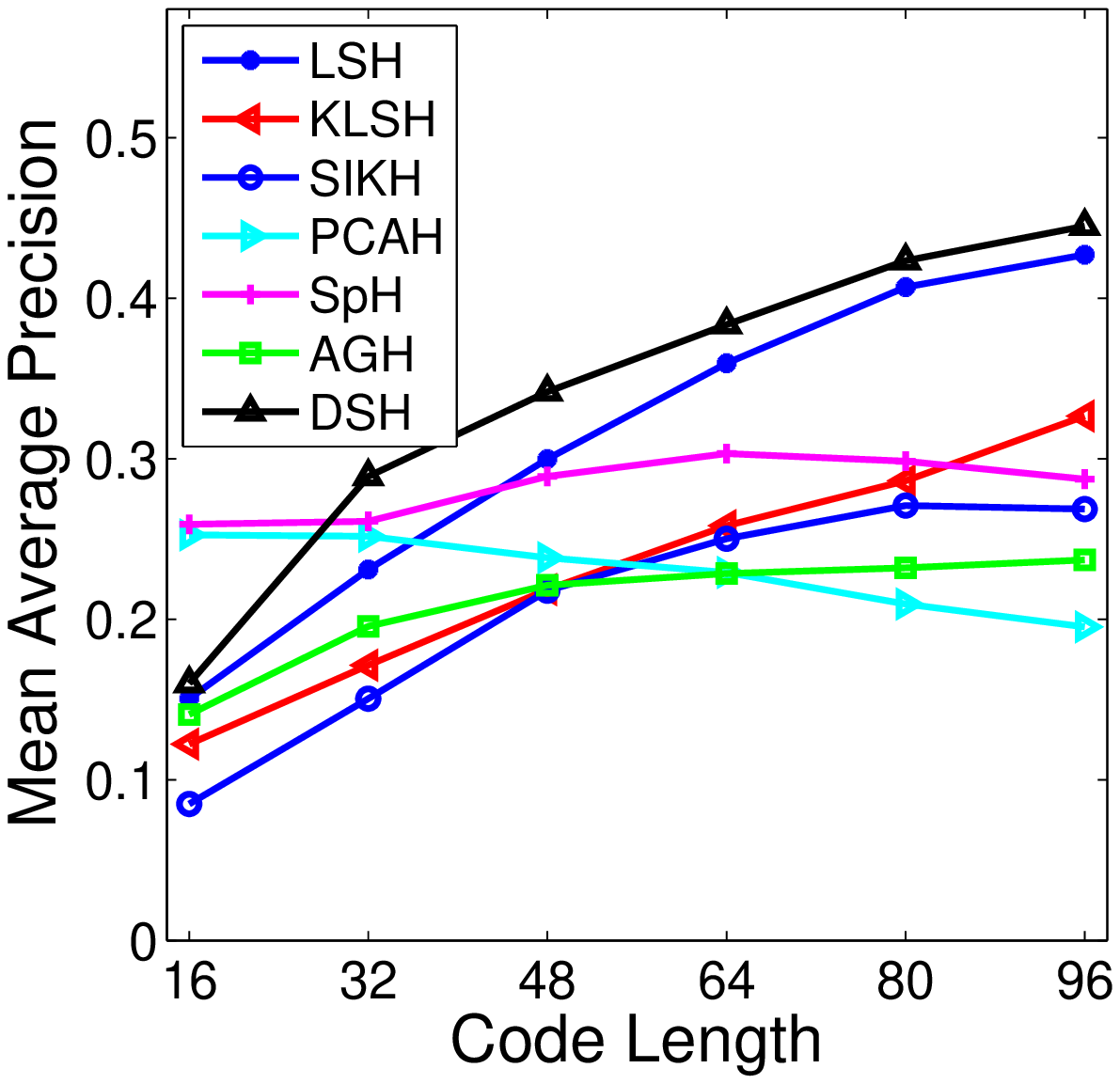}}
\caption{The Mean Average Precision of all the algorithms on the three data sets.} \label{fig:2}
\end{center}
\end{figure*}

\subsection{Computational Complexity Analysis}

Given $n$ data points with the dimensionality $d$, the computational complexity of DSH in the training stage is as follows:
\begin{enumerate}
\item $O(\alpha Lpnd)$: $k$-means with $p$ iterations to generate $\alpha L$ groups (Step 1 in Alg. \ref{alg:1}).
\item $O((\alpha L)^{2}(d+r))$: Find all the $r$-adjacent groups (Step 2 in Alg. \ref{alg:1}).
\item $O(\alpha Lrd)$: For each pair of adjacent groups, generate the projection and the intercept (Step 3 in Alg. \ref{alg:1}).
\item Compute the entropy for all the candidate projections needs $O((\alpha L)^{2}dr)$ (Step 4 in Alg. \ref{alg:1}).
\item The top $L$ projections can be found within $O(\alpha Lr\log(\alpha Lr))$. The binary codes for data points can be obtained in $O(Lnd)$ (Step 5 in Alg. \ref{alg:1}).
\end{enumerate}
Considering $\alpha Lr \ll n$, the overall computational complexity of DSH training is dominated by the $k$-means clustering step which is $O(\alpha Lpnd)$. It is clear that DSH scales linearly with respect to the number of samples in the database.

In the testing stage, given a query point, DSH needs $O(Ld)$ to compress the query point into a binary code, which is the same as the complexity of Locality Sensitive Hashing.

\section{Experiment}

In this section, we evaluate our DSH algorithm on the high dimensional nearest neighbor search problem. Three large scale real-world data sets are used in our experiments.
\begin{itemize}
\item \textbf{GIST1M}: It contains one million GIST features and each feature is represented by a 960-dim vector. This data set is publicly available\footnote{http://corpus-texmex.irisa.fr}.
\item \textbf{Flickr1M}: We collect one million images from the Flickr and use a feature extraction code\footnote{http://www.vision.ee.ethz.ch/$\sim$zhuji/felib.html} to extract a GIST feature for each image. Each image is represented by a 512-dim GIST feature vector. This data set is publicly available\footnote{http://www.cad.zju.edu.cn/home/dengcai/Data/NNSData.html}.
\item \textbf{SIFT1M}: It contains one million SIFT features and each feature is represented by a 128-dim vector. This data set is publicly available\footnote{http://corpus-texmex.irisa.fr}.
\end{itemize}

For each data set, we randomly select 1k data points as the queries and use the remaining to form the gallery database. We use the same criterion as in~\cite{wang2010sequential,xu2011complementary}, that a returned point is considered to be a true neighbor if it lies in the top 2 percentile points closest (measured by the Euclidian distance in the original space) to the query.
For each query, all the data points in the database are ranked according to their Hamming distances to the query. We evaluate the retrieval results by the Mean Average Precision (MAP) and  the precision-recall curve \cite{wang2010sequential}. In addition, we also report the training time and the testing time (the average time used for each query) for all the methods.

\begin{figure*}[t!]
\begin{center}
\includegraphics[width=0.65\columnwidth]{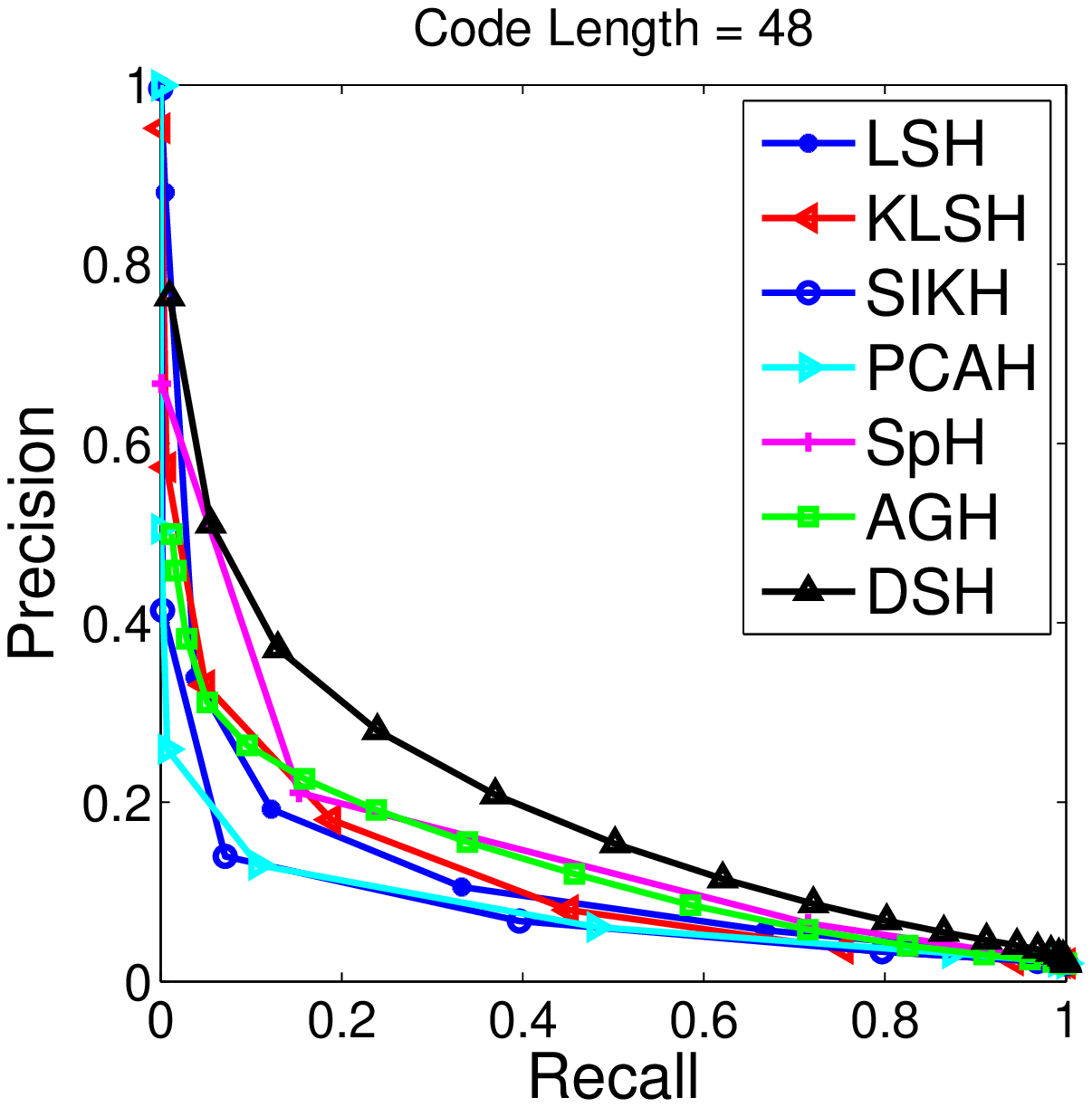}
\includegraphics[width=0.65\columnwidth]{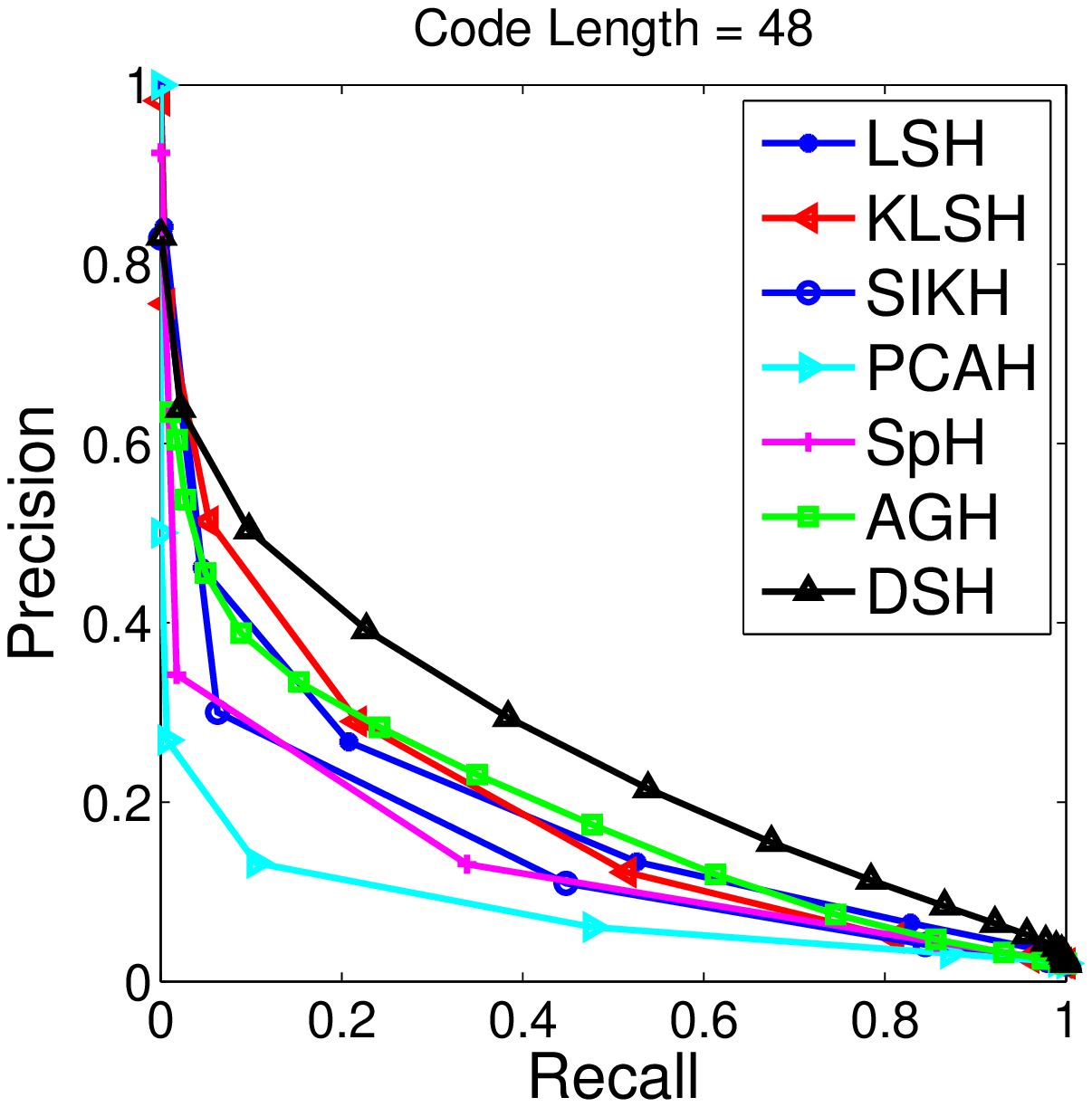}
\includegraphics[width=0.65\columnwidth]{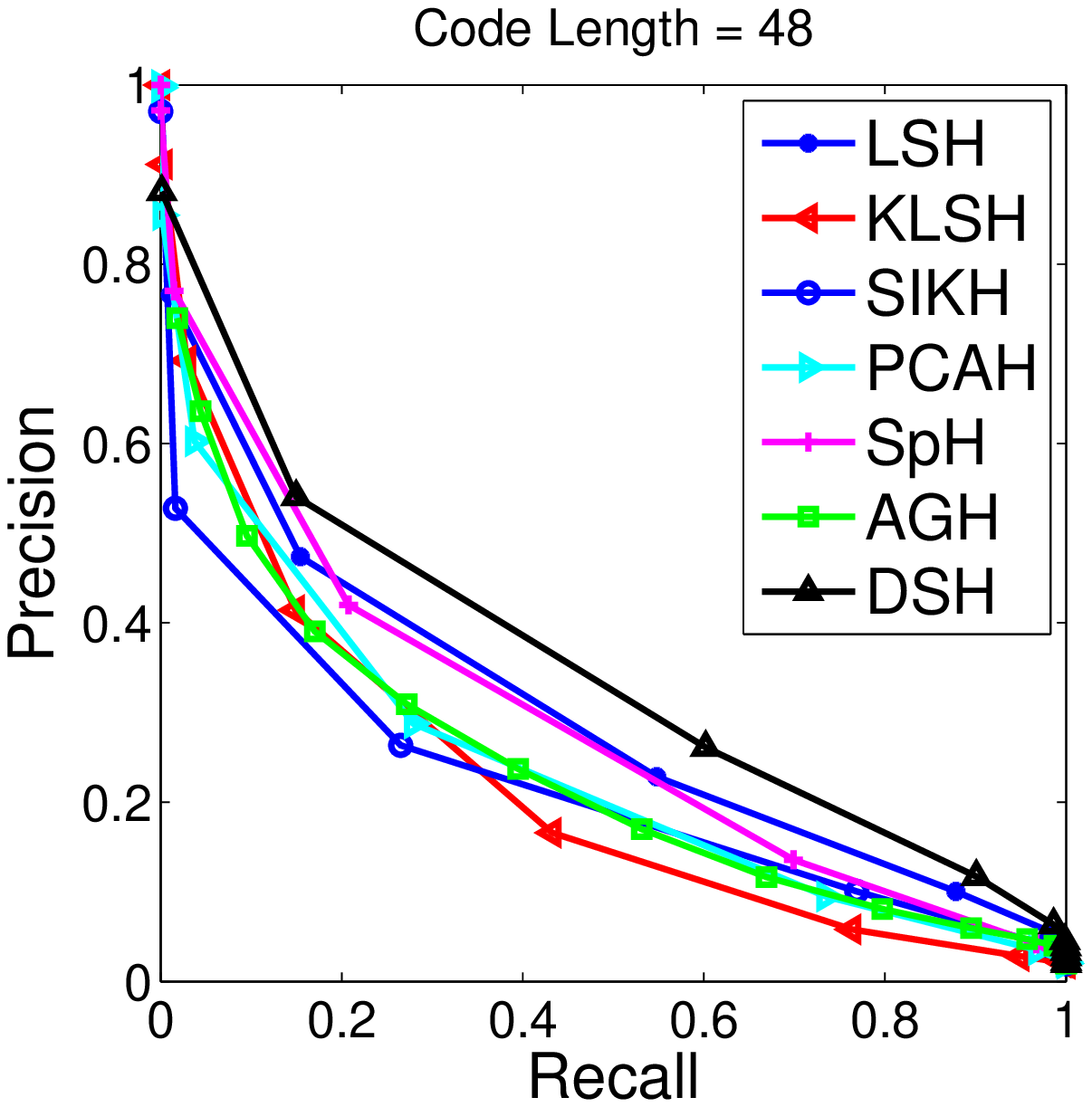}
\subfigure[GIST1M]{
\includegraphics[width=0.65\columnwidth]{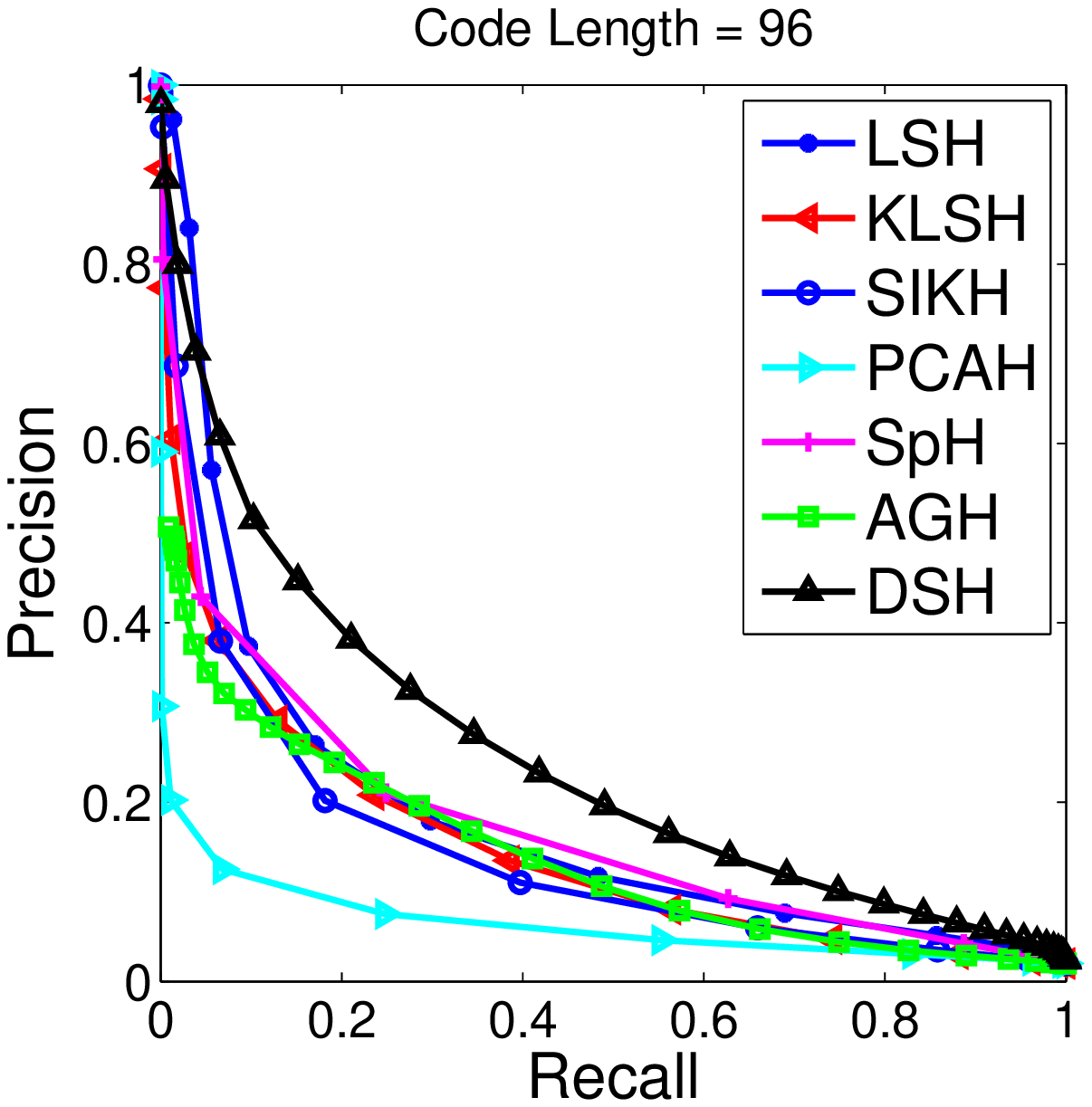}}
\subfigure[Flickr1M]{
\includegraphics[width=0.65\columnwidth]{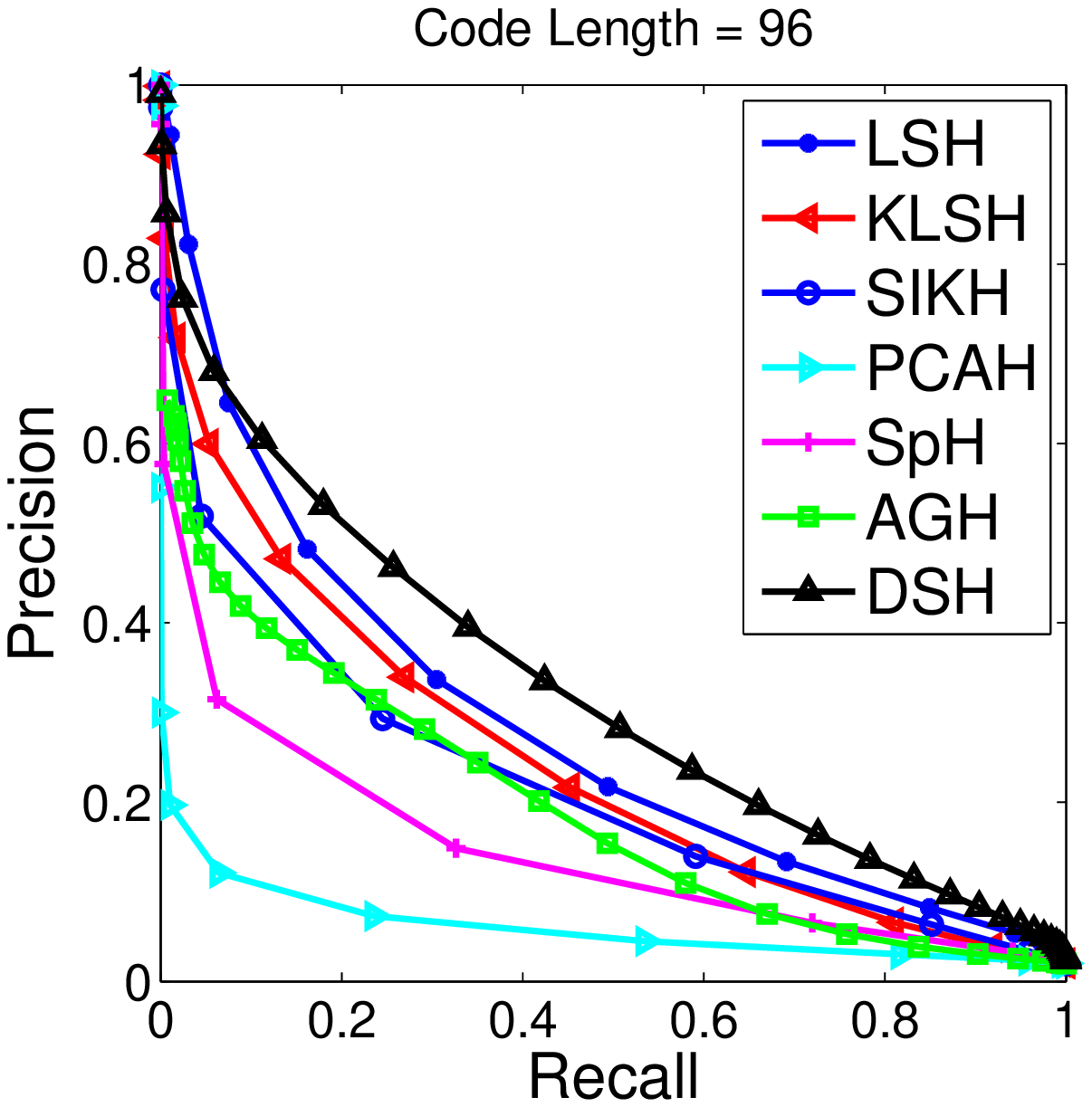}}
\subfigure[SIFT1M]{
\includegraphics[width=0.65\columnwidth]{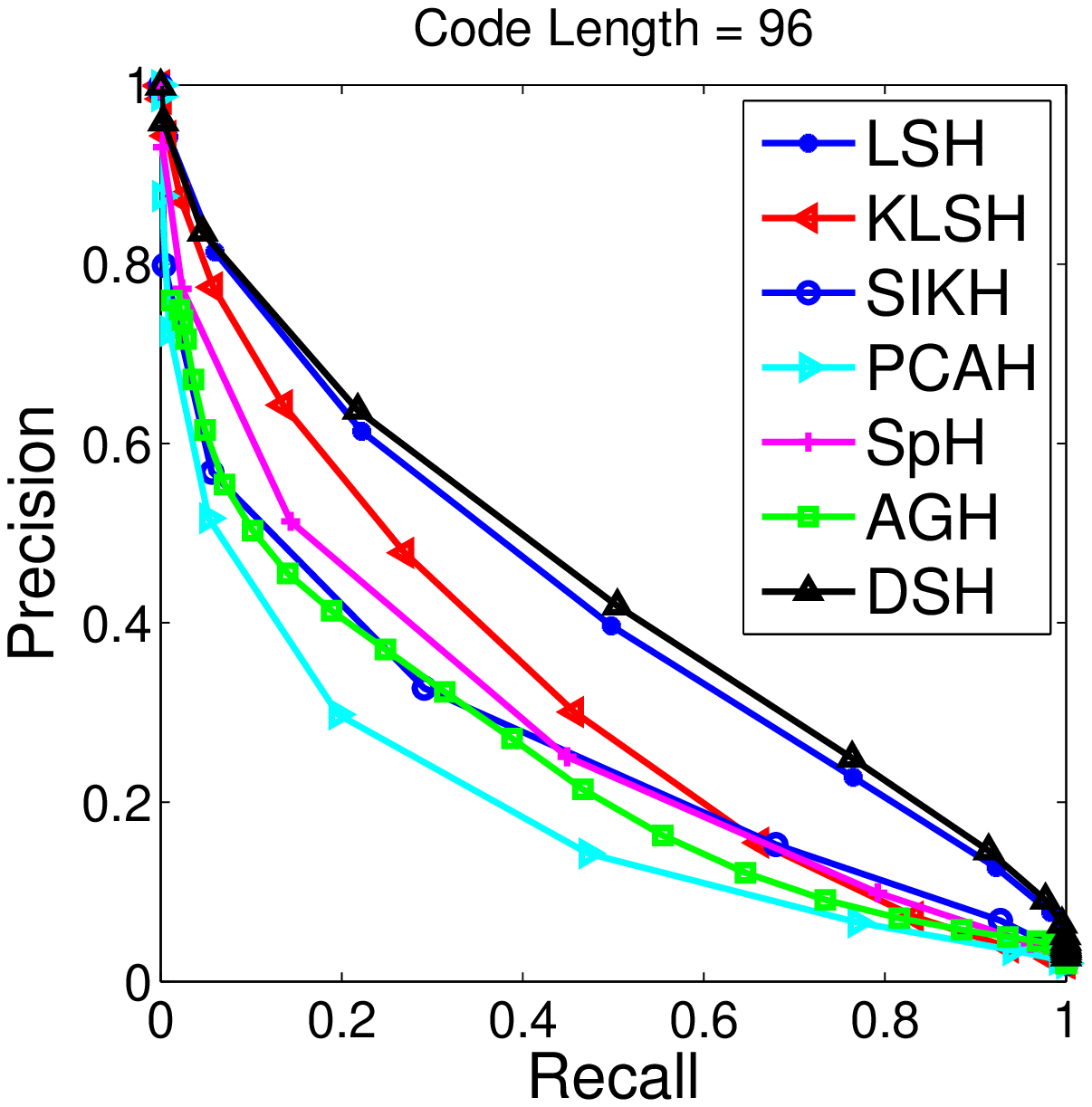}}
\caption{The precision-recall curves of all algorithms on three data sets for the codes of 48 bits and 96 bits.} \label{fig:3}
\end{center}
\vskip -20pt
\end{figure*}

\subsection{Compared Algorithms}
Seven state-of-the-art hashing algorithms for high dimensional nearest neighbors search are compared as follows:
\begin{itemize}
\item Locality Sensitive Hashing (\textbf{LSH})~\cite{charikar2002similarity}, which is based on the random projection. The projective vectors are randomly sampling from a $p$-stable distribution (\eg, Gaussian). We implement the algorithm by ourselves and make it publicly available\footnote{http://www.cad.zju.edu.cn/home/dengcai/Data/DSH.html}.
\item Kernelized Locality Sensitive Hashing (\textbf{KLSH})~\cite{kulis2009kernelized},  which generalizes the LSH method to the kernel space. We use the code provided by the authors\footnote{http://www.cse.ohio-state.edu/$\sim$kulis/klsh/klsh.htm}. \nop{ of KLSH provide the code  To run KLSH, we use the Gaussian kernel and sample 300 training points to form the empirical kernel map. The bandwidth of Gaussian kernel is set to 0.3.}
\item Shift-Invariant Kernel Hashing (\textbf{SIKH})~\cite{raginsky2009locality}, which is a distribution-free method based on the random features mapping for approximating shift-invariant kernels. The code is also publicly available\footnote{http://www.unc.edu/$\sim$yunchao/itq.htm}.
\item Principle Component Analysis Hashing (\textbf{PCAH})~\cite{wang2006annosearch}, which directly uses the top principal directions as the projective vectors to obtain the binary codes. The implementation of PCA is publicly available\footnote{http://www.cad.zju.edu.cn/home/dengcai/Data/DimensionReduction.html}.
\item Spectral Hashing (\textbf{SpH})~\cite{weiss2008spectral}, which is based on quantizing the values of analytical eigenfunctions computed along PCA directions of the data. We use the code provided by the authors\footnote{http://www.cs.huji.ac.il/$\sim$yweiss/SpectralHashing/}.
\item Anchor Graph Hashing (\textbf{AGH})~\cite{wei2011hashing}, which constructs an anchor graph to speed up the spectral analysis procedure. AGH with two-layer is used in our comparison for its superior performance over AGH with one-layer \cite{wei2011hashing}. We use the code provided by the authors\footnote{http://www.ee.columbia.edu/$\sim$wliu/} and the number of anchors is set to be 300 and the number of nearest neighbors is set to be 2 as suggested in \cite{wei2011hashing}.
\item Density Sensitive Hashing ({\bf DSH}), which is the method introduced in this paper. For the purpose of reproducibility, we also make the code publicly available\footnote{http://www.cad.zju.edu.cn/home/dengcai/Data/DSH.html}.  There are three parameters. We empirically set $p=3$ (the number of iterations in $k$-means), $\alpha = 1.5$ (controlling the groups number), $r=3$ (for $r$-adjacent groups). A detailed analysis on the parameter selection will be provided later.
\end{itemize}

It is important to note that LSH, KLSH and SIKH are random projection based methods, while PCAH, SpH and AGH are learning based methods. Our DSH can be regarded as a combination of these two directions.

\begin{table*}
\caption{Training and testing time of all algorithms on GIST1M.} \label{tb:1}
\begin{center}
\begin{tabular}{l||cccc|rrrr}
\hline
& \multicolumn{4}{|c|}{Training Time (s)} & \multicolumn{4}{|c}{Test Time (s)}\\
\hline
Method & $L=16$ & $L=32$ & $L=64$ & $L=96$ & \multicolumn{1}{c}{$L=16$} & \multicolumn{1}{c}{$L=32$} & \multicolumn{1}{c}{$L=64$} & \multicolumn{1}{c}{$L=96$}\\
\hline
LSH~\cite{charikar2002similarity} & 0.4 & 1.0 & 2.3 & 2.6 & $1.2 \times 10^{-6}$ & $2.6 \times 10^{-6}$ & $5.8 \times 10^{-6}$ & $7.1 \times 10^{-6}$\\
KLSH~\cite{kulis2009kernelized} & 27.4 & 27.7 & 27.9 & 28.3 & $30.0 \times 10^{-6}$ & $32.3 \times 10^{-6}$ & $34.7 \times 10^{-6}$ & $36.5 \times 10^{-6}$\\
SIKH~\cite{raginsky2009locality} & 1.3 & 2.5 & 3.8 & 5.2 & $3.9 \times 10^{-6}$ & $6.3 \times 10^{-6}$ & $10.5 \times 10^{-6}$ & $15.9 \times 10^{-6}$\\
\hline
PCAH~\cite{wang2006annosearch} & 31.3 & 57.2 & 60.3 & 75.0 & $1.2 \times 10^{-6}$ & $2.7 \times 10^{-6}$ & $5.6 \times 10^{-6}$ & $7.2 \times 10^{-6}$\\
SpH~\cite{weiss2008spectral} & 42.5 & 77.8 & 125.3 & 239.8 & $23.9 \times 10^{-6}$ & $42.1 \times 10^{-6}$ & $93.4 \times 10^{-6}$ & $270.1 \times 10^{-6}$\\
AGH~\cite{wei2011hashing} & 340.8 & 344.7 & 349.8 & 356.0 & $33.3 \times 10^{-6}$ & $52.6 \times 10^{-6}$ & $71.2 \times 10^{-6}$ & $191.3 \times 10^{-6}$\\
\hline
DSH & 33.1 & 45.9 & 56.5 & 63.6 & $1.3 \times 10^{-6}$ & $2.7 \times 10^{-6}$ & $5.8 \times 10^{-6}$ & $7.1 \times 10^{-6}$\\
\hline
\end{tabular}
\end{center}
\vskip -7pt
\end{table*}

\begin{table*}
\caption{Training and testing time of all algorithms on Flickr1M.} \label{tb:2}
\begin{center}
\begin{tabular}{l||cccc|rrrr}
\hline
& \multicolumn{4}{|c|}{Training Time (s)} & \multicolumn{4}{|c}{Test Time (s)}\\
\hline
Method & $L=16$ & $L=32$ & $L=64$ & $L=96$ & \multicolumn{1}{c}{$L=16$} & \multicolumn{1}{c}{$L=32$} & \multicolumn{1}{c}{$L=64$} & \multicolumn{1}{c}{$L=96$}\\
\hline
LSH~\cite{charikar2002similarity} & 0.3 & 0.8 &  1.5 & 1.8 & $0.9 \times 10^{-6}$ & $2.0 \times 10^{-6}$ & $2.8 \times 10^{-6}$ & $4.6 \times 10^{-6}$\\
KLSH~\cite{kulis2009kernelized} & 18.2 & 18.5 &  18.9 & 19.4 & $15.2 \times 10^{-6}$ & $18.9 \times 10^{-6}$ & $22.8 \times 10^{-6}$ & $25.1 \times 10^{-6}$\\
SIKH~\cite{raginsky2009locality} & 1.1 & 2.0 & 2.8 & 4.3 & $2.8 \times 10^{-6}$ & $3.8 \times 10^{-6}$ & $9.1 \times 10^{-6}$ & $12.3 \times 10^{-6}$\\
\hline
PCAH~\cite{wang2006annosearch} & 16.7 & 29.4 & 31.4 & 33.7 & $1.1 \times 10^{-6}$ & $2.1 \times 10^{-6}$ & $2.9 \times 10^{-6}$ & $4.9 \times 10^{-6}$\\
SpH~\cite{weiss2008spectral} & 22.3 & 45.6 & 106.2 & 205.5 & $16.7 \times 10^{-6}$ & $38.5 \times 10^{-6}$ & $88.6 \times 10^{-6}$ & $251.6 \times 10^{-6}$\\
AGH~\cite{wei2011hashing} & 232.9 & 247.9 & 257.4 & 268.1 & $28.2 \times 10^{-6}$ & $42.2 \times 10^{-6}$ & $52.2 \times 10^{-6}$ & $155.3 \times 10^{-6}$\\
\hline
DSH & 17.4 & 29.3 & 35.8 & 45.9 & $0.9 \times 10^{-6}$ & $2.1 \times 10^{-6}$  & $2.8 \times 10^{-6}$ & $4.6 \times 10^{-6}$\\
\hline
\end{tabular}
\end{center}
\vskip -7pt
\end{table*}

\begin{table*}
\caption{Training and testing time of all algorithms on SIFT1M.} \label{tb:3}
\begin{center}
\begin{tabular}{l||cccc|rrrr}
\hline
& \multicolumn{4}{|c|}{Training Time (s)} & \multicolumn{4}{|c}{Test Time (s)}\\
\hline
Method & $L=16$ & $L=32$ & $L=64$ & $L=96$ & \multicolumn{1}{c}{$L=16$} & \multicolumn{1}{c}{$L=32$} & \multicolumn{1}{c}{$L=64$} & \multicolumn{1}{c}{$L=96$}\\
\hline
LSH~\cite{charikar2002similarity} & 0.1 & 0.3 & 0.6 & 0.8 & $0.4 \times 10^{-6}$ & $1.1 \times 10^{-6}$ & $1.8 \times 10^{-6}$ & $2.4 \times 10^{-6}$\\
KLSH~\cite{kulis2009kernelized} & 10.2 & 10.4 & 10.8 & 11.2 & $12.2 \times 10^{-6}$ & $13.1 \times 10^{-6}$ & $13.8 \times 10^{-6}$ & $15.7 \times 10^{-6}$\\
SIKH~\cite{raginsky2009locality} & 0.5 & 1.1 & 2.3 & 3.5 & $0.9 \times 10^{-6}$ & $2.3 \times 10^{-6}$ & $6.3 \times 10^{-6}$ & $7.0 \times 10^{-6}$\\
\hline
PCAH~\cite{wang2006annosearch} & 3.9 & 6.5 & 7.5 & 7.8 & $0.5 \times 10^{-6}$ & $1.3 \times 10^{-6}$ & $2.0 \times 10^{-6}$ & $2.5 \times 10^{-6}$\\
SpH~\cite{weiss2008spectral} & 11.4 & 28.1 & 92.7 & 189.1 & $11.8 \times 10^{-6}$ & $33.3 \times 10^{-6}$ & $77.1 \times 10^{-6}$ & $230.9 \times 10^{-6}$\\
AGH~\cite{wei2011hashing} & 135.2 & 142.5 & 148.1 & 155.1 & $15.3 \times 10^{-6}$ & $23.9 \times 10^{-6}$ & $31.2 \times 10^{-6}$ & $57.1 \times 10^{-6}$\\
\hline
DSH & 8.4 & 12.2 & 15.5 & 20.1 & $0.5 \times 10^{-6}$ & $1.2 \times 10^{-6}$ & $1.9 \times 10^{-6}$ & $2.6 \times 10^{-6}$\\
\hline
\end{tabular}
\end{center}
\vskip -7pt
\end{table*}

\subsection{Experimental Results}

Figure~\ref{fig:2} shows the MAP curves of all the algorithms on the GIST1M, Flickr1M and SIFT1M data sets, respectively. We can see that the three random projection based methods (LSH, KLSH and SIKH) have a low MAP when the code length is short. As the code length increases, the performances of all the three methods consistently increases. On the other hand, the learning based methods (PCAH, SpH and AGH) have a high MAP when the code length is short. However, they fail to make significant improvements as the code length increases. Particulary, the performance of PCAH decreases as the code length increases. This is consistent with previous work \cite{gong2011iterative,wang2010sequential} and is probably because that most of the data variance is contained in the top few principal directions so that the later bits are calculated using the low-variance projections, leading to the poorly discriminative codes \cite{wang2010sequential}. By utilizing the geometric structure of the data to guide the projections selection, our DSH successfully combines the advantages of both random projection based methods and the learning based methods. As a result, DSH achieves a satisfied performance on the three data sets and almost outperforms its competitors for all code lengths. It is interesting to see that the performance improvements of DSH over other methods on GIST1M and Flickr1M are larger than that on SIFT1M. Since the dimensions of the data in GIST1M ($960$-d) and Flickr1M ($512$-d) are much larger than that in SIFT1M ($128$-d), this suggests that our DSH method are particularly suitable for high dimensional situations. Figure~\ref{fig:3} presents the precision-recall curves of all the algorithms on three data sets with the codes of 48 bits and 96 bits.

Table~\ref{tb:1}, \ref{tb:2} and \ref{tb:3} show both the training and testing time for different algorithms on three data sets, respectively. We can clearly see that both the training and testing time of all the methods decrease as the dimension of the data decreases.
Considering the training time, the three random projection based algorithms are relatively efficient, especially for LSH and SIKH. KLSH needs to compute a sampled kernel matrix which slows down its computation. The three learning based algorithms are relatively slow, for exploring the data structure. Our DSH is also fast. Although it is slower than the three random projection based algorithms, it is significantly faster than SpH and AGH.
Considering the testing time, LSH, PCAH and our DSH are the most efficient methods.
All of them simply need a matrix multiplication and a thresholding to obtain the binary codes. SpH consumes much longer time than other methods as the code length increases since it needs to compute the analytical eigenfunctions involving the calculation of trigonometric functions.

\begin{figure*}[t]
\begin{center}
\subfigure[GIST1M]{
\includegraphics[width=\TriQuaterFigureWidth]{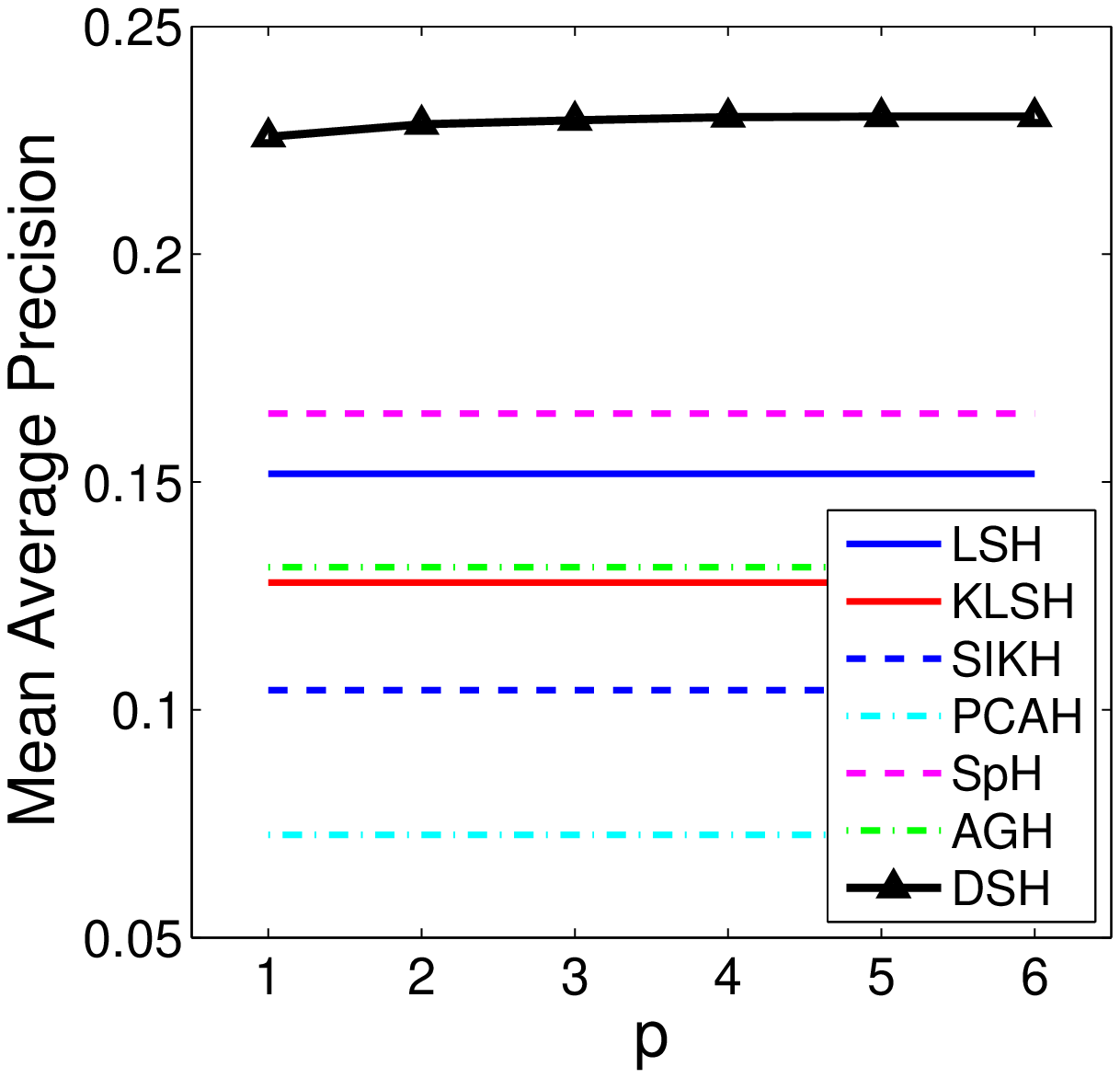}}
\subfigure[Flickr1M]{
\includegraphics[width=\TriQuaterFigureWidth]{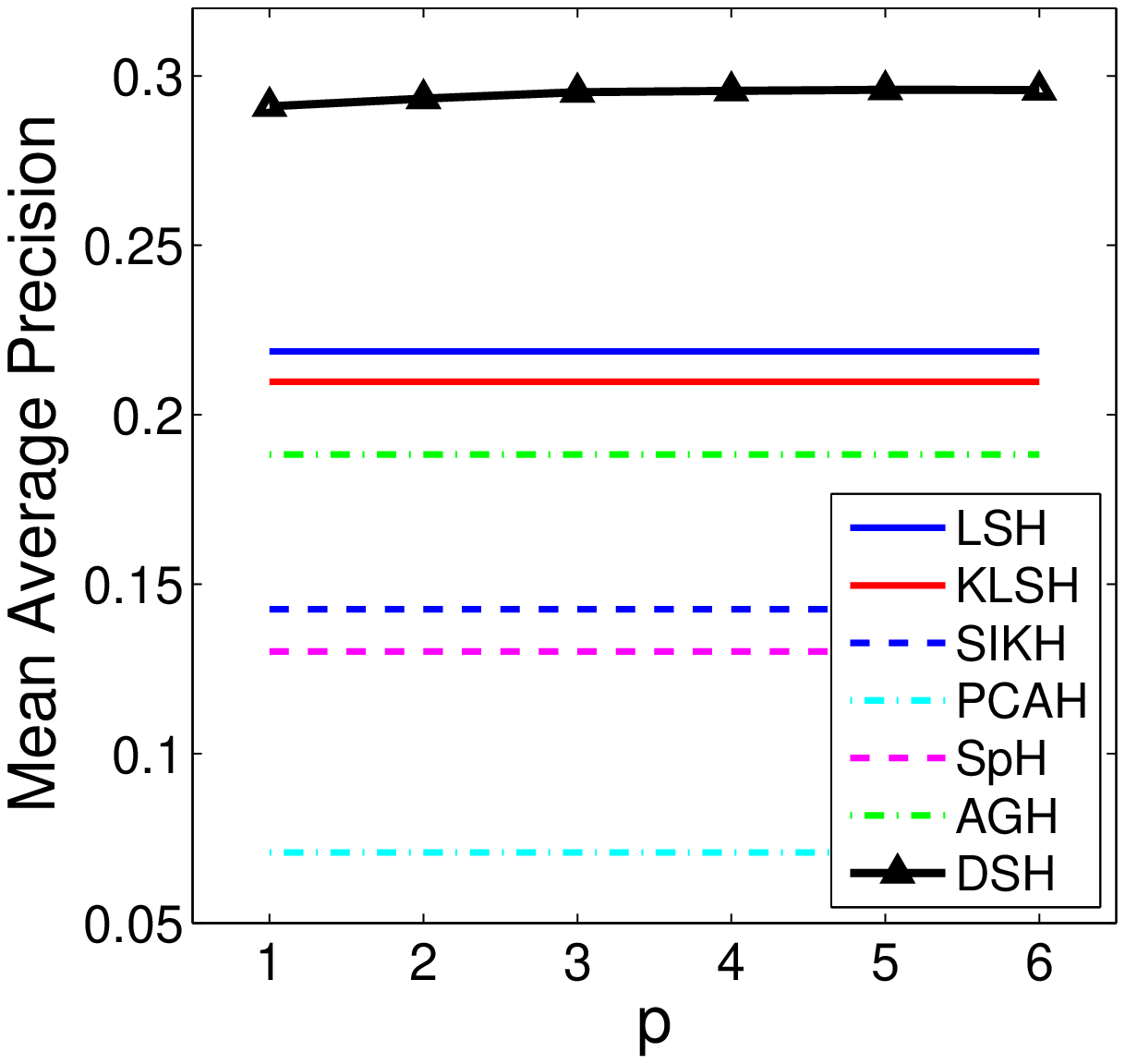}}
\subfigure[SIFT1M]{
\includegraphics[width=\TriQuaterFigureWidth]{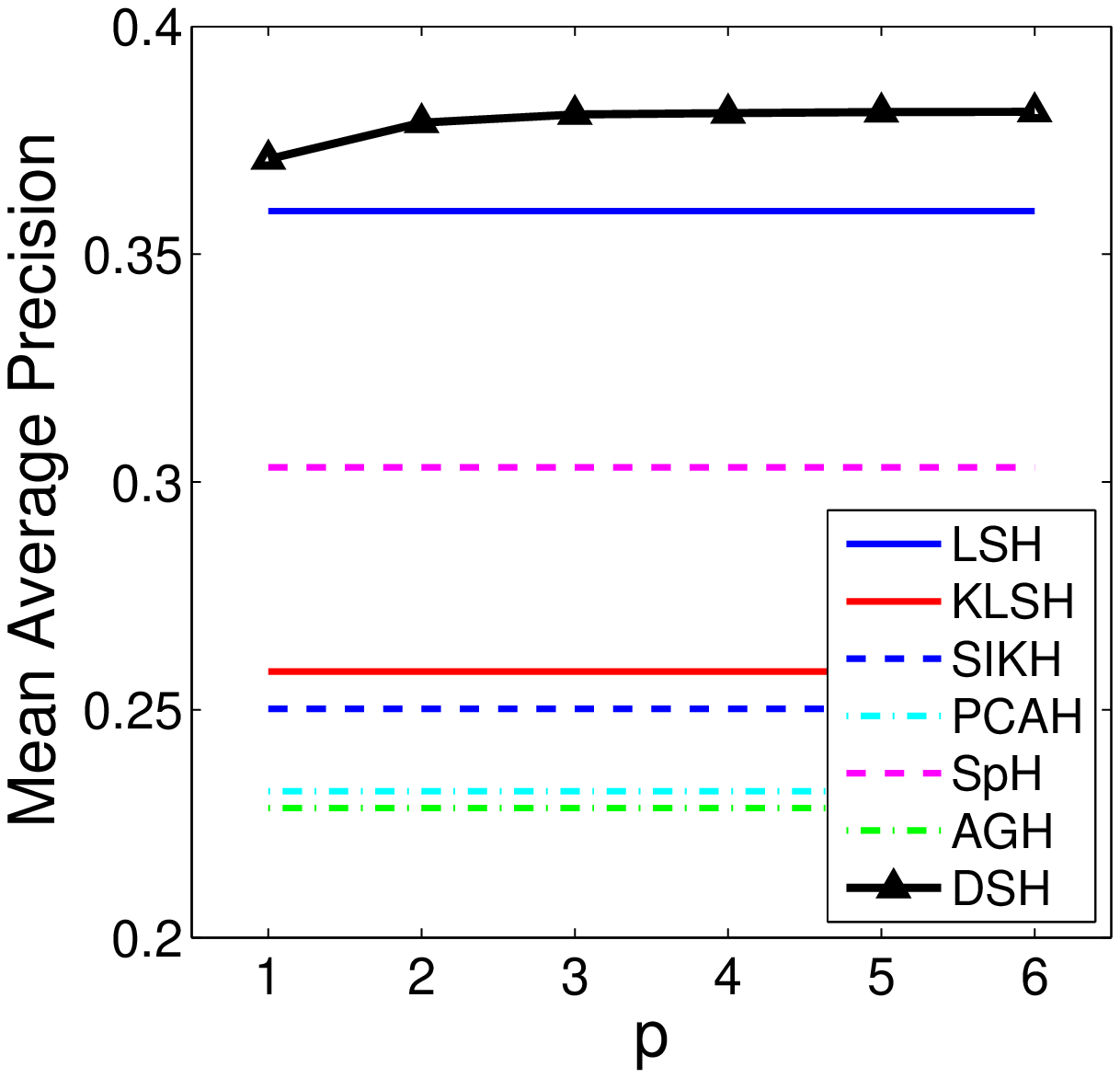}}
\caption{The performance of DSH vs. the number of iterations of $k$-means ($p$) at 64 bits.} \label{fig:4}

\tabcaption{Training time (s) of DSH vs. the number of iterations of $k$-means ($p$) at 64 bits.} \label{tb:4}
\begin{tabular}{c|cccccc}
\hline
Data Set & $p=1$ & $p=2$ & $p=3$ & $p=4$ & $p=5$ & $p=6$\\
\hline
GIST1M & 18.8 & 37.2 & 56.5 & 76.2 & 94.1 & 111.7\\
Flickr1M & 11.7 & 23.5 & 35.8 & 48.1 & 62.6 & 76.4\\
SIFT1M & 4.8 & 9.1 & 15.5 & 21.2 & 25.5 & 31.9\\
\hline
\end{tabular}
\end{center}
\end{figure*}

\begin{figure*}[t]
\begin{center}
\subfigure[GIST1M]{
\includegraphics[width=\TriQuaterFigureWidth]{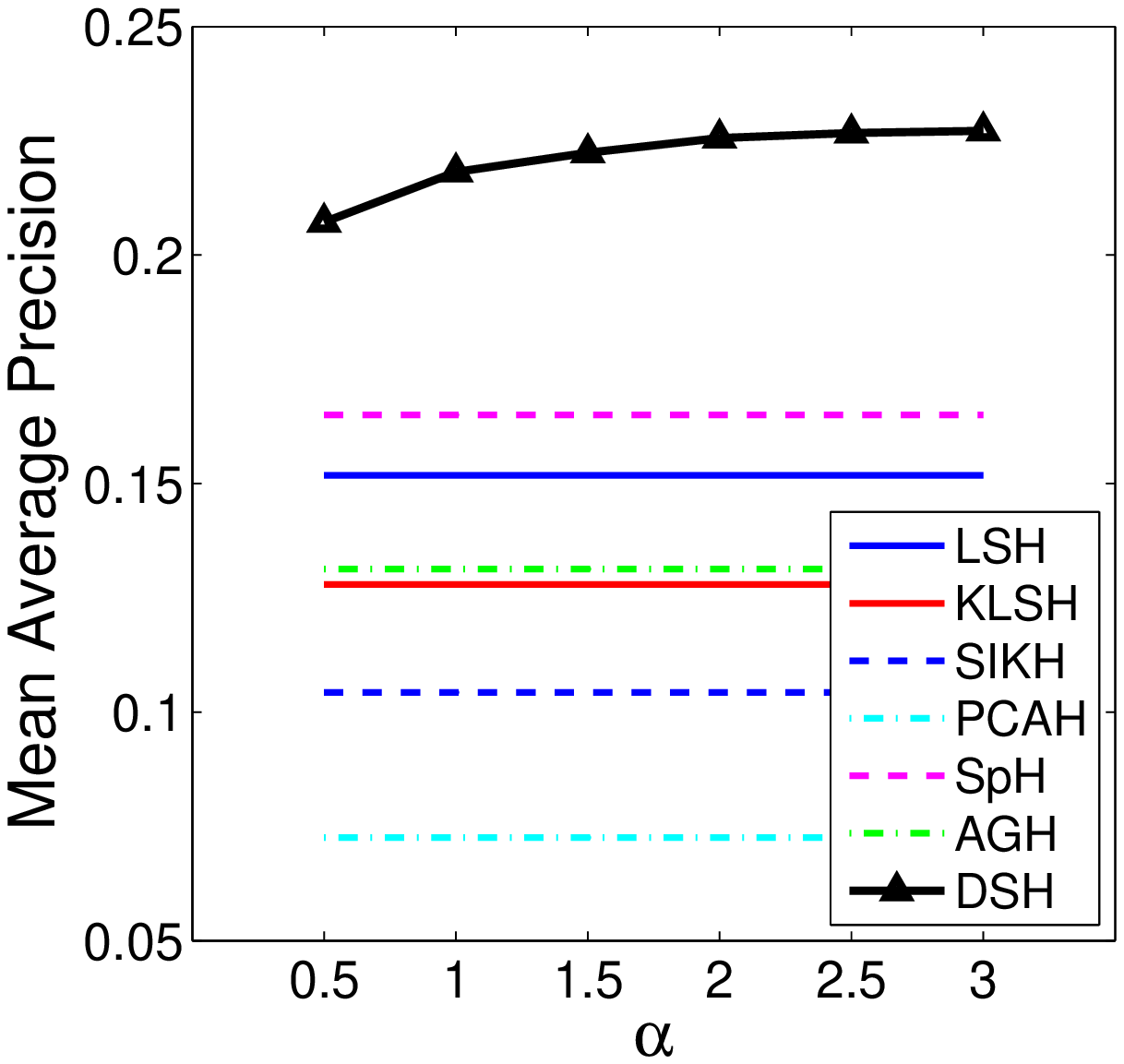}}
\subfigure[Flickr1M]{
\includegraphics[width=\TriQuaterFigureWidth]{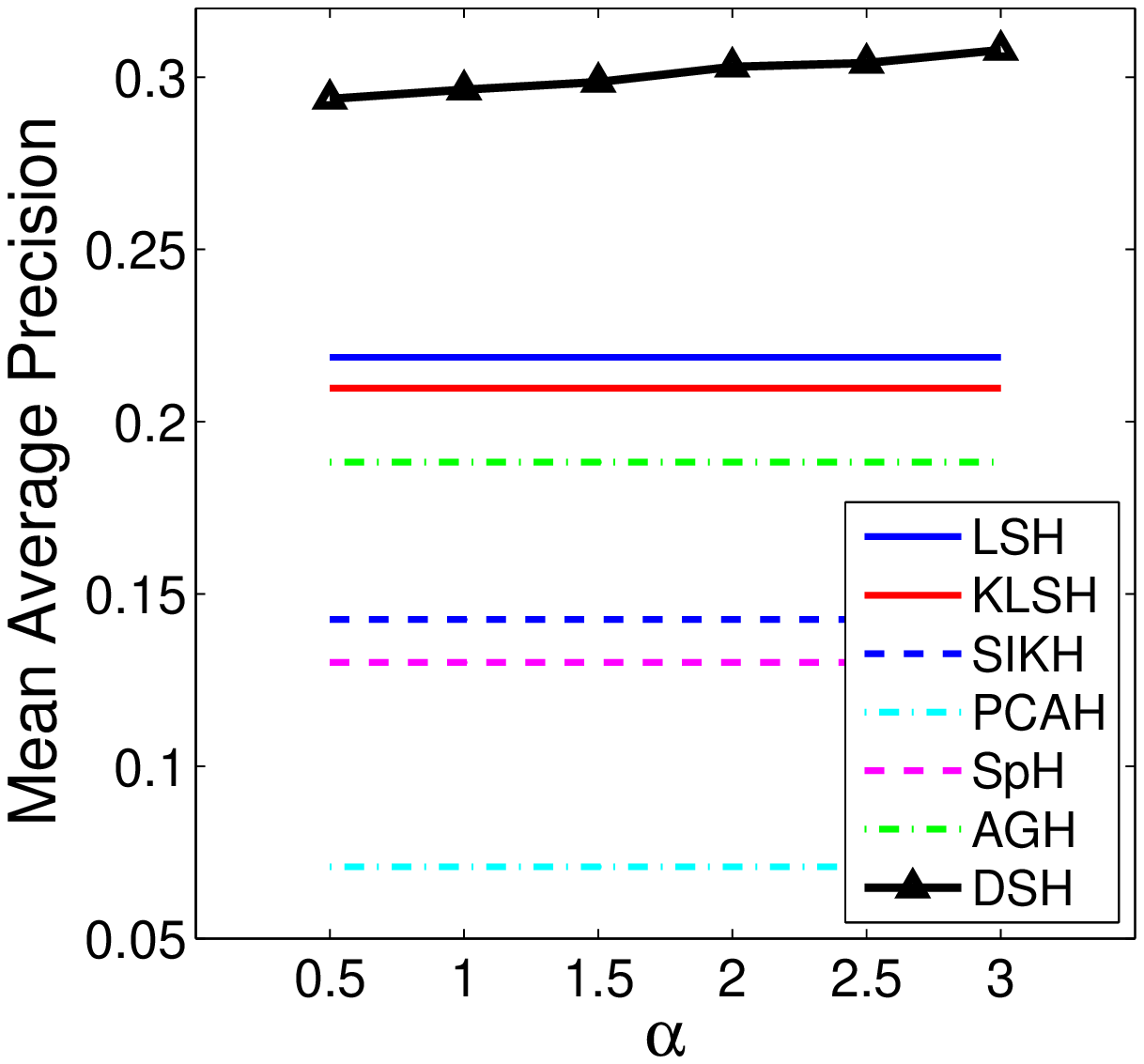}}
\subfigure[SIFT1M]{
\includegraphics[width=\TriQuaterFigureWidth]{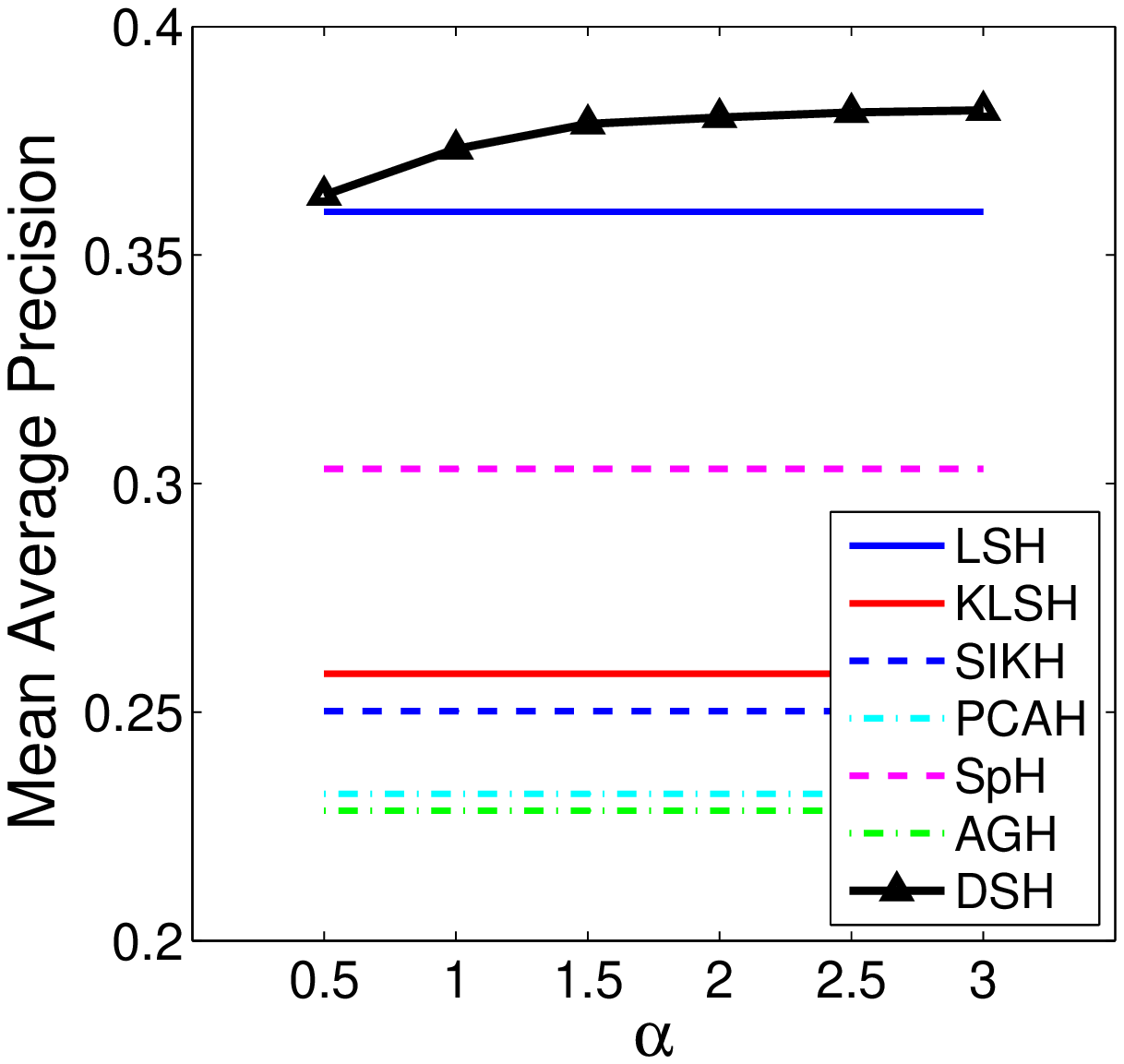}}
\caption{The performance of DSH vs. the parameter $\alpha$ (controlling the number of groups) at 64 bits.} \label{fig:6}

\tabcaption{Training time (s) of DSH vs. the parameter $\alpha$ (controlling the number of groups) at 64 bits.} \label{tb:5}
\begin{tabular}{c|cccccc}
\hline
Data Set & $\alpha=0.5$ & $\alpha=1.0$ & $\alpha=1.5$ & $\alpha=2.0$ & $\alpha=2.5$ & $\alpha=3.0$\\
\hline
GIST1M & 48.4 & 52.9 & 56.5 & 68.4 & 74.6 & 83.5\\
Flickr1M & 21.8 & 25.3 & 35.8 & 46.7 & 57.3 & 68.2\\
SIFT1M & 9.8 & 11.2 & 15.5 & 21.3 & 28.2 & 37.9\\
\hline
\end{tabular}

\end{center}
\end{figure*}

\subsection{Parameter Selection}

Our DSH has three parameters: $p$ (the number of iterations in $k$-means), $\alpha$ (the parameter controlling the groups number) and $r$ (the parameter for $r$-adjacent groups). In this subsection, we discuss how the performance of DSH will be influenced by these three parameters. We learn 64-bits hashing codes and the default setting for these parameters is $p=3$, $\alpha=1.5$ and $r=3$. When we study the impact of one parameter, the other parameters are fixed as the default.

Figure~\ref{fig:4} and Table~\ref{tb:4} show how the performance of DSH varies as the number of iterations in $k$-means varies. As the
number of iterations increases, it is reasonable to see that both the MAP
and the learning time of DSH increase. On all the three data sets,
3 iterations in $k$-means are enough for achieving reasonably good MAP.

Figure~\ref{fig:6} and Table~\ref{tb:5} show how the performance of DSH varies as $\alpha$ changes (the groups number generated by $k$-means changes). As we can see, as $\alpha$ becomes larger (the groups number increases), both the MAP and learning time of DSH increase.
Setting $\alpha = 1.5$ is a reasonable balance considering both the accuracy and the efficiency.

\begin{figure*}[t]
\begin{center}
\subfigure[GIST1M]{
\includegraphics[width=\TriQuaterFigureWidth]{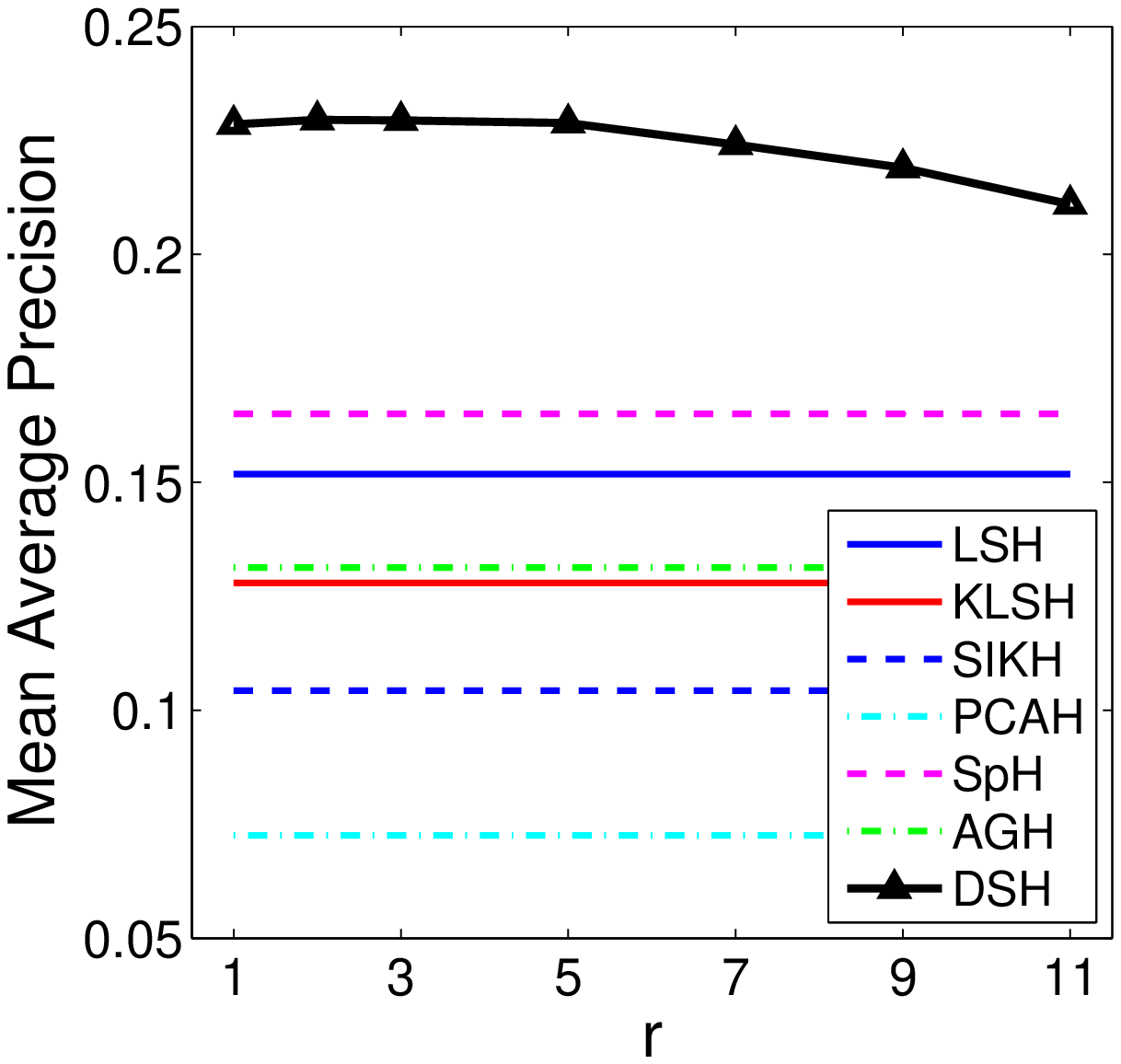}}
\subfigure[Flickr1M]{
\includegraphics[width=\TriQuaterFigureWidth]{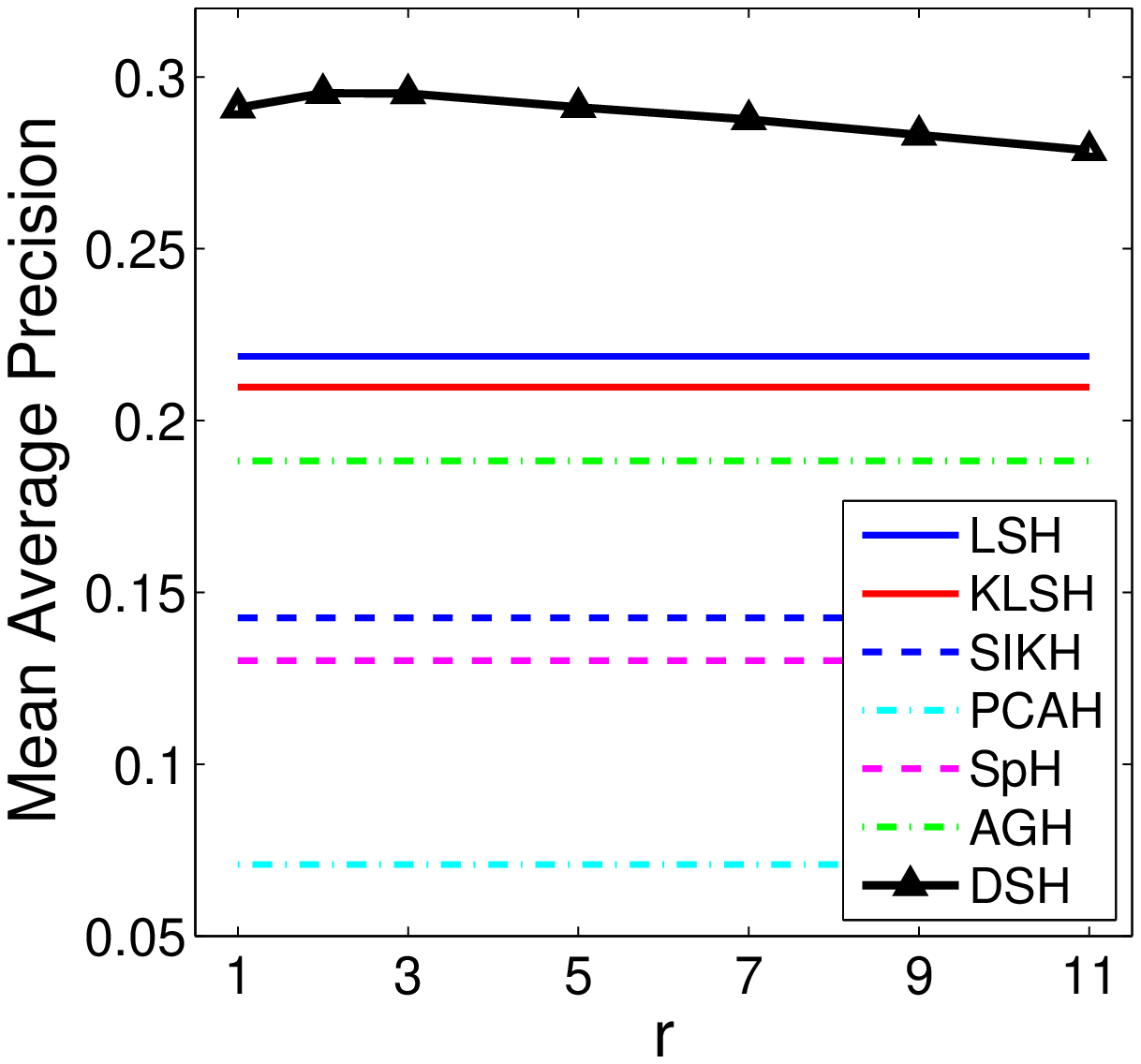}}
\subfigure[SIFT1M]{
\includegraphics[width=\TriQuaterFigureWidth]{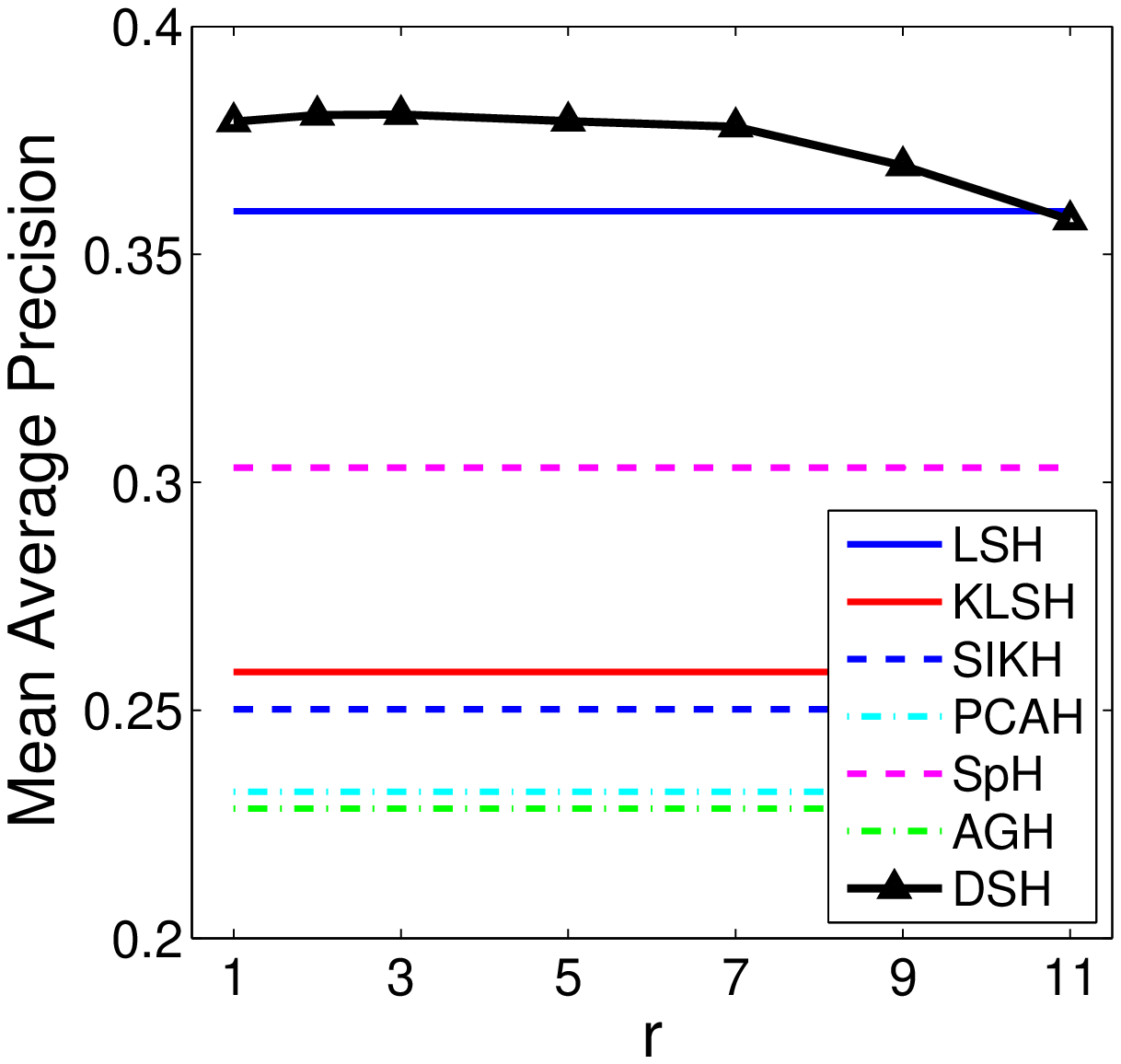}}
\caption{The performance of DSH vs. the parameter $r$ (for $r$-adjacent groups) at 64 bits. } \label{fig:5}
\end{center}
\end{figure*}

Figure~\ref{fig:5} shows the performance of DSH varies as $r$ ($r$-adjacent groups) changes.
DSH achieves stable and consistent good performance as $r$ is less than 5. As $r$ becomes larger, DSH generates more projections which are used to separate two far away groups.
These projections are usually less critical and redundant. Thus, the performance of DSH decreases.

\section{Conclusion}

In this paper, we have developed a novel hashing algorithm, called {\em Density Sensitive Hashing} (DSH), for high dimensional nearest neighbors search. Different from those random projection based hashing approaches, \eg, Locality Sensitive Hashing, DSH uses the geometric structure of the data to guide the projections selection. As a result, DSH can generate hashing codes with more discriminating power.
Empirical studies on three large data sets show that the proposed algorithm scales well to data size and significantly outperforms the state-of-the-art hashing methods in terms of retrieval accuracy.


\end{document}